\pdfoutput=1

\documentclass[11pt,twoside,a4paper,cmspaper,final,collab]{cms-tdr}
\def\svnVersion{249994}\def\cmsCernNo{2014-032}\def\cmsCernDate{\today}\def\cmsMessage{Published in the Journal of High Energy Physics as \href{http://dx.doi.org/10.1007/JHEP06(2014)090}{\doi{10.1007/JHEP06(2014)090.}}}

\begin{document}\cmsNoteHeader{TOP-12-038}

\hyphenation{had-ron-i-za-tion}
\hyphenation{cal-or-i-me-ter}
\hyphenation{de-vices}

\RCS$Revision: 234163 $
\RCS$HeadURL: svn+ssh://svn.cern.ch/reps/tdr2/papers/TOP-12-038/trunk/TOP-12-038.tex $
\RCS$Id: TOP-12-038.tex 234163 2014-03-28 12:28:28Z alverson $
\newlength{\cmsFigWidth}\setlength{\cmsFigWidth}{0.48\textwidth}
\newcommand{\tch}{\ensuremath{t\text{-ch.}}}
\renewcommand{\thy}{\ensuremath{\,\text{(theo.)}}\xspace}
\providecommand{\exper}{\ensuremath{\,\text{(exp.)}}\xspace}
\newcommand{\Vtb}{\ensuremath{{V_{\mathrm{tb}}}}\xspace}
\newcommand{\vtb}{\Vtb}
\newcommand{\Vtd}{\ensuremath{{V_{\mathrm{td}}}}\xspace}
\newcommand{\vtd}{\Vtd}
\newcommand{\Vts}{\ensuremath{{V_{\mathrm{ts}}}}\xspace}
\newcommand{\vts}{\Vts}
\newcommand{\absvtb}{\ensuremath{{\abs{V_{\mathrm{tb}}}}}\xspace}
\newcommand{\absvtd}{\ensuremath{{\abs{V_{\mathrm{td}}}}}\xspace}
\newcommand{\absvts}{\ensuremath{{\abs{V_{\mathrm{ts}}}}}\xspace}
\newcommand{\absVtb}{\absvtb}
\newcommand{\absVtd}{\absvtd}
\newcommand{\absVts}{\absvts}
\newcommand{\flv}{\ensuremath{f_\mathrm{{Lv}}}\xspace}
\newcommand{\absflvvtb}{\ensuremath{\abs{f_{\mathrm{Lv}}V_{\mathrm{tb}}}}\xspace}
\newcommand{\etalj}{\ensuremath{\eta_{j'}}\xspace}
\newcommand{\absetalj}{\ensuremath{\abs{\eta_{j'}}}\xspace}
\newcommand{\mpt}{\ensuremath{{\textbf{p}\!\!\!/}_{\mathrm{\textbf{\!T}}}}\xspace}
\newcommand{\mpx}{\ensuremath{{p\!\!\!/}_{\!\!x}}\xspace}
\newcommand{\mpy}{\ensuremath{{p\!\!\!/}_{\!\!y}}\xspace}
\newcommand{\PFrelIso}{\ensuremath{I_{\text{rel}}}\xspace}
\newcommand{\PFRelIso}{\ensuremath{I_{\text{rel}}}\xspace}
\newcommand{\PFrelIsoRho}{\ensuremath{I_{\text{rel}}}\xspace}
\newcommand{\pfPhotonIso}{\ensuremath{I^{\gamma}}\xspace}
\newcommand{\pfChargedHadronIso}{\ensuremath{I^{\text{ch.\,h}}}\xspace}
\newcommand{\pfNeutralHadronIso}{\ensuremath{I^{\text{n.\,h}}}\xspace}
\newcommand{\pfPU}{\ensuremath{I^{\mathrm{PU}}\xspace}}
\newcommand{\sumPUPT}{\ensuremath{\sum p_{\mathrm{T}}^{\mathrm{PU}}}\xspace}
\newcommand{\sumPUPt}{\sumPUPT}
\newcommand{\rhoEnergy}{\ensuremath{\rho \times A}}
\newcommand{\mTW}{\ensuremath{m_{\mathrm{T}}}\xspace}
\newcommand{\mtw}{\ensuremath{m_{\mathrm{T}}}\xspace}
\newcommand{\mW}{\ensuremath{m_{\PW}}\xspace}
\newcommand{\mT}{\ensuremath{m_{\mathrm{T}}}\xspace}
\newcommand{\qcd}{\ensuremath{\mathrm{QCD}}\xspace}
\newcommand{\QCD}{\ensuremath{\mathrm{QCD}~\text{multijet}}\xspace}
\newcommand{\ttot}{\ensuremath{\sigma_{\tch}}\xspace}
\newcommand{\ttotth}{\ensuremath{\sigma^{\text{theo.}}_{\tch}}\xspace}
\newcommand{\tplus}{\ensuremath{\sigma_{\tch}(\cPqt)}\xspace}
\newcommand{\tminus}{\ensuremath{\sigma_{\tch}(\cPaqt}\xspace}
\newcommand{\wjets}{\ensuremath{\PW\text{+jets}}\xspace}
\newcommand{\WJets}{\wjets}
\newcommand{\wzjets}{\ensuremath{\PW/\cPZ\text{+jets}}\xspace}
\newcommand{\wzcjets}{\ensuremath{\PW/\cPZ\text{+c\,jets}}\xspace}
\newcommand{\WZjets}{\wzjets}
\newcommand{\vjets}{\wzjets}
\newcommand{\VJets}{\wzjets}
\newcommand{\Vjets}{\wzjets}
\newcommand{\vcjets}{\wzcjets}
\newcommand{\Vcjets}{\wzcjets}
\newcommand{\zjets}{\ensuremath{\cPZ\text{+jets}}\xspace}
\newcommand{\ZJets}{\zjets}
\newcommand{\wzheavy}{\ensuremath{\PW/\cPZ\text{+heavy}}\xspace}
\newcommand{\wzhf}{\ensuremath{\PW/cPZ\text{+heavy flavours}}\xspace}
\newcommand{\ttoleptoncascade}{\ensuremath{\cPqt\to \PW\cPqb\to \cPqb\ell\nu}\xspace}
\newcommand{\ttotaucascade}{\ensuremath{\cPqt \to \PW\cPqb\to \cPqb\tau\nu}\xspace}
\renewcommand{\tt}{\ttbar}
\newcommand{\mt}{\ensuremath{m_{\ell\nu \cPqb}}\xspace}
\newcommand{\cme}{centre-of-mass energy\xspace}
\newcommand{\topMass}{\mt}
\newcommand{\tq}{\cPqt\xspace}
\newcommand{\tbar}{\cPaqt\xspace}
\newcommand{\tqbar}{\tbar}
\newcommand{\xsectheor}{87.2\xspace}
\newcommand{\xsectoptheor}{56.4\xspace}
\newcommand{\xsecantitoptheor}{30.7\xspace}
\newcommand{\ratiotheor}{1.8\xspace}
\newcommand{\totmueles}{0.97\xspace}
\newcommand{\totmuelescorr}{0.96\xspace}
\newcommand{\totmuelesxsec}{84.6\xspace}
\newcommand{\totmuelescorrxsec}{83.6\xspace}
\newcommand{\totmueleewkmu}{1.08\xspace}
\newcommand{\totmueleewkmucorr}{1.02\xspace}
\newcommand{\totmueleewkele}{1.19\xspace}
\newcommand{\totmueleewkelecorr}{1.09\xspace}
\newcommand{\totmueletop}{0.99\xspace}
\newcommand{\totmueletopcorr}{1.01\xspace}
\newcommand{\totmuelestatss}{0.03\xspace}
\newcommand{\totmuelestatsscorr}{0.03\xspace}
\newcommand{\totmuelestatssxsec}{2.3\xspace}
\newcommand{\totmuelestatsscorrxsec}{2.3\xspace}
\newcommand{\totmuelestatsewkmu}{0.09\xspace}
\newcommand{\totmuelestatsewkmucorr}{0.00\xspace}
\newcommand{\totmuelestatsewkele}{0.12\xspace}
\newcommand{\totmuelestatsewkelecorr}{0.00\xspace}
\newcommand{\totmuelestatstop}{0.04\xspace}
\newcommand{\totmuelestatstopcorr}{0.00\xspace}
\newcommand{\totmuelesystss}{0.09\xspace}
\newcommand{\totmuelesystsscorr}{0.09\xspace}
\newcommand{\totmuelesystssxsec}{7.5\xspace}
\newcommand{\totmuelesystsscorrxsec}{7.4\xspace}
\newcommand{\totmuelesystsewkmu}{0.00\xspace}
\newcommand{\totmuelesystsewkmucorr}{0.00\xspace}
\newcommand{\totmuelesystsewkele}{0.00\xspace}
\newcommand{\totmuelesystsewkelecorr}{0.00\xspace}
\newcommand{\totmuelesyststop}{0.00\xspace}
\newcommand{\totmuelesyststopcorr}{0.00\xspace}
\newcommand{\totmuelelumis}{0.03\xspace}
\newcommand{\totmuelelumiscorr}{0.02\xspace}
\newcommand{\totmuelelumisxsec}{2.20\xspace}
\newcommand{\totmuelelumiscorrxsec}{2.17\xspace}
\newcommand{\totmuelelumiewkmu}{0.00\xspace}
\newcommand{\totmuelelumiewkmucorr}{0.00\xspace}
\newcommand{\totmuelelumiewkele}{0.00\xspace}
\newcommand{\totmuelelumiewkelecorr}{0.00\xspace}
\newcommand{\totmuelelumitop}{0.00\xspace}
\newcommand{\totmuelelumitopcorr}{0.00\xspace}
\newcommand{\totmuelesystsnolumis}{0.08\xspace}
\newcommand{\totmuelesystsnolumiscorr}{0.08\xspace}
\newcommand{\totmuelesystsnolumisxsec}{7.19\xspace}
\newcommand{\totmuelesystsnolumiscorrxsec}{7.11\xspace}
\newcommand{\totmuelesystsnolumiewkmu}{0.00\xspace}
\newcommand{\totmuelesystsnolumiewkmucorr}{0.00\xspace}
\newcommand{\totmuelesystsnolumiewkele}{0.00\xspace}
\newcommand{\totmuelesystsnolumiewkelecorr}{0.00\xspace}
\newcommand{\totmuelesystsnolumitop}{0.00\xspace}
\newcommand{\totmuelesystsnolumitopcorr}{0.00\xspace}
\newcommand{\totmus}{0.96\xspace}
\newcommand{\totmuscorr}{0.96\xspace}
\newcommand{\totmusxsec}{84.02\xspace}
\newcommand{\totmuscorrxsec}{83.86\xspace}
\newcommand{\totmuewkmu}{1.09\xspace}
\newcommand{\totmuewkmucorr}{1.04\xspace}
\newcommand{\totmuewkele}{0.00\xspace}
\newcommand{\totmuewkelecorr}{0.00\xspace}
\newcommand{\totmutop}{0.99\xspace}
\newcommand{\totmutopcorr}{1.01\xspace}
\newcommand{\totmustatss}{0.03\xspace}
\newcommand{\totmustatsscorr}{0.03\xspace}
\newcommand{\totmustatssxsec}{2.92\xspace}
\newcommand{\totmustatsscorrxsec}{2.91\xspace}
\newcommand{\totmustatsewkmu}{0.12\xspace}
\newcommand{\totmustatsewkmucorr}{0.00\xspace}
\newcommand{\totmustatsewkele}{0.00\xspace}
\newcommand{\totmustatsewkelecorr}{0.00\xspace}
\newcommand{\totmustatstop}{0.06\xspace}
\newcommand{\totmustatstopcorr}{0.00\xspace}
\newcommand{\totmusystss}{0.09\xspace}
\newcommand{\totmusystsscorr}{0.09\xspace}
\newcommand{\totmusystssxsec}{7.56\xspace}
\newcommand{\totmusystsscorrxsec}{7.55\xspace}
\newcommand{\totmusystsewkmu}{0.00\xspace}
\newcommand{\totmusystsewkmucorr}{0.00\xspace}
\newcommand{\totmusystsewkele}{0.00\xspace}
\newcommand{\totmusystsewkelecorr}{0.00\xspace}
\newcommand{\totmusyststop}{0.00\xspace}
\newcommand{\totmusyststopcorr}{0.00\xspace}
\newcommand{\totmulumis}{0.03\xspace}
\newcommand{\totmulumiscorr}{0.03\xspace}
\newcommand{\totmulumisxsec}{2.18\xspace}
\newcommand{\totmulumiscorrxsec}{2.18\xspace}
\newcommand{\totmulumiewkmu}{0.00\xspace}
\newcommand{\totmulumiewkmucorr}{0.00\xspace}
\newcommand{\totmulumiewkele}{0.00\xspace}
\newcommand{\totmulumiewkelecorr}{0.00\xspace}
\newcommand{\totmulumitop}{0.00\xspace}
\newcommand{\totmulumitopcorr}{0.00\xspace}
\newcommand{\totmusystsnolumis}{0.08\xspace}
\newcommand{\totmusystsnolumiscorr}{0.08\xspace}
\newcommand{\totmusystsnolumisxsec}{7.24\xspace}
\newcommand{\totmusystsnolumiscorrxsec}{7.23\xspace}
\newcommand{\totmusystsnolumiewkmu}{0.00\xspace}
\newcommand{\totmusystsnolumiewkmucorr}{0.00\xspace}
\newcommand{\totmusystsnolumiewkele}{0.00\xspace}
\newcommand{\totmusystsnolumiewkelecorr}{0.00\xspace}
\newcommand{\totmusystsnolumitop}{0.00\xspace}
\newcommand{\totmusystsnolumitopcorr}{0.00\xspace}
\newcommand{\toteles}{0.98\xspace}
\newcommand{\totelescorr}{0.95\xspace}
\newcommand{\totelesxsec}{85.79\xspace}
\newcommand{\totelescorrxsec}{82.90\xspace}
\newcommand{\toteleewkmu}{0.00\xspace}
\newcommand{\toteleewkmucorr}{0.00\xspace}
\newcommand{\toteleewkele}{1.18\xspace}
\newcommand{\toteleewkelecorr}{1.08\xspace}
\newcommand{\toteletop}{0.99\xspace}
\newcommand{\toteletopcorr}{1.02\xspace}
\newcommand{\totelestatss}{0.04\xspace}
\newcommand{\totelestatsscorr}{0.04\xspace}
\newcommand{\totelestatssxsec}{3.70\xspace}
\newcommand{\totelestatsscorrxsec}{3.57\xspace}
\newcommand{\totelestatsewkmu}{0.00\xspace}
\newcommand{\totelestatsewkmucorr}{0.00\xspace}
\newcommand{\totelestatsewkele}{0.15\xspace}
\newcommand{\totelestatsewkelecorr}{0.00\xspace}
\newcommand{\totelestatstop}{0.05\xspace}
\newcommand{\totelestatstopcorr}{0.00\xspace}
\newcommand{\totelesystss}{0.10\xspace}
\newcommand{\totelesystsscorr}{0.10\xspace}
\newcommand{\totelesystssxsec}{9.14\xspace}
\newcommand{\totelesystsscorrxsec}{8.83\xspace}
\newcommand{\totelesystsewkmu}{0.00\xspace}
\newcommand{\totelesystsewkmucorr}{0.00\xspace}
\newcommand{\totelesystsewkele}{0.00\xspace}
\newcommand{\totelesystsewkelecorr}{0.00\xspace}
\newcommand{\totelesyststop}{0.00\xspace}
\newcommand{\totelesyststopcorr}{0.00\xspace}
\newcommand{\totelelumis}{0.03\xspace}
\newcommand{\totelelumiscorr}{0.02\xspace}
\newcommand{\totelelumisxsec}{2.23\xspace}
\newcommand{\totelelumiscorrxsec}{2.16\xspace}
\newcommand{\totelelumiewkmu}{0.00\xspace}
\newcommand{\totelelumiewkmucorr}{0.00\xspace}
\newcommand{\totelelumiewkele}{0.00\xspace}
\newcommand{\totelelumiewkelecorr}{0.00\xspace}
\newcommand{\totelelumitop}{0.00\xspace}
\newcommand{\totelelumitopcorr}{0.00\xspace}
\newcommand{\totelesystsnolumis}{0.10\xspace}
\newcommand{\totelesystsnolumiscorr}{0.10\xspace}
\newcommand{\totelesystsnolumisxsec}{8.86\xspace}
\newcommand{\totelesystsnolumiscorrxsec}{8.56\xspace}
\newcommand{\totelesystsnolumiewkmu}{0.00\xspace}
\newcommand{\totelesystsnolumiewkmucorr}{0.00\xspace}
\newcommand{\totelesystsnolumiewkele}{0.00\xspace}
\newcommand{\totelesystsnolumiewkelecorr}{0.00\xspace}
\newcommand{\totelesystsnolumitop}{0.00\xspace}
\newcommand{\totelesystsnolumitopcorr}{0.00\xspace}
\newcommand{\chmuelesminus}{0.98\xspace}
\newcommand{\chmuelesminuscorr}{0.90\xspace}
\newcommand{\chmuelesminusxsec}{30.23\xspace}
\newcommand{\chmuelesminuscorrxsec}{27.6\xspace}
\newcommand{\chmuelesplus}{1.01\xspace}
\newcommand{\chmuelespluscorr}{0.95\xspace}
\newcommand{\chmuelesplusxsec}{56.72\xspace}
\newcommand{\chmuelespluscorrxsec}{53.8\xspace}
\newcommand{\chmuelesratio}{1.02\xspace}
\newcommand{\chmuelesratiocorr}{1.06\xspace}
\newcommand{\chmuelesratioR}{1.88\xspace}
\newcommand{\chmuelesratioRcorr}{1.95\xspace}
\newcommand{\chmueleewkmuminus}{0.92\xspace}
\newcommand{\chmueleewkeleminus}{0.97\xspace}
\newcommand{\chmueleewkratio}{1.04\xspace}
\newcommand{\chmueletop}{1.06\xspace}
\newcommand{\chmuelestatssminus}{0.05\xspace}
\newcommand{\chmuelestatssminuscorr}{0.04\xspace}
\newcommand{\chmuelestatssminusxsec}{1.47\xspace}
\newcommand{\chmuelestatssminuscorrxsec}{1.3\xspace}
\newcommand{\chmuelestatssplus}{0.03\xspace}
\newcommand{\chmuelestatsspluscorr}{0.03\xspace}
\newcommand{\chmuelestatssplusxsec}{1.6\xspace}
\newcommand{\chmuelestatsspluscorrxsec}{1.5\xspace}
\newcommand{\chmuelestatssratio}{0.05\xspace}
\newcommand{\chmuelestatssratiocorr}{0.05\xspace}
\newcommand{\chmuelestatssratioR}{0.09\xspace}
\newcommand{\chmuelestatssratioRcorr}{0.10\xspace}
\newcommand{\chmuelestatsewkmuminus}{0.09\xspace}
\newcommand{\chmuelestatsewkeleminus}{0.13\xspace}
\newcommand{\chmuelestatsewkratio}{0.04\xspace}
\newcommand{\chmuelestatstop}{0.04\xspace}
\newcommand{\chmuelesystssminus}{0.13\xspace}
\newcommand{\chmuelesystssminuscorr}{0.12\xspace}
\newcommand{\chmuelesystssminusxsec}{4.05\xspace}
\newcommand{\chmuelesystssminuscorrxsec}{3.7\xspace}
\newcommand{\chmuelesystssplus}{0.08\xspace}
\newcommand{\chmuelesystsspluscorr}{0.08\xspace}
\newcommand{\chmuelesystssplusxsec}{4.66\xspace}
\newcommand{\chmuelesystsspluscorrxsec}{4.4\xspace}
\newcommand{\chmuelesystssratio}{0.10\xspace}
\newcommand{\chmuelesystssratiocorr}{0.10\xspace}
\newcommand{\chmuelesystssratioR}{0.18\xspace}
\newcommand{\chmuelesystssratioRcorr}{0.19\xspace}
\newcommand{\chmuelesystsewkmuminus}{0.00\xspace}
\newcommand{\chmuelesystsewkeleminus}{0.00\xspace}
\newcommand{\chmuelesystsewkratio}{0.00\xspace}
\newcommand{\chmuelesyststop}{0.00\xspace}
\newcommand{\chmuelelumisminus}{0.03\xspace}
\newcommand{\chmuelelumisminuscorr}{0.02\xspace}
\newcommand{\chmuelelumisminusxsec}{0.79\xspace}
\newcommand{\chmuelelumisminuscorrxsec}{0.72\xspace}
\newcommand{\chmuelelumisplus}{0.03\xspace}
\newcommand{\chmuelelumispluscorr}{0.02\xspace}
\newcommand{\chmuelelumisplusxsec}{1.47\xspace}
\newcommand{\chmuelelumispluscorrxsec}{1.4\xspace}
\newcommand{\chmuelelumisratio}{0.00\xspace}
\newcommand{\chmuelelumisratiocorr}{0.00\xspace}
\newcommand{\chmuelelumisratioR}{0.00\xspace}
\newcommand{\chmuelelumisratioRcorr}{0.00\xspace}
\newcommand{\chmuelelumiewkmuminus}{0.00\xspace}
\newcommand{\chmuelelumiewkeleminus}{0.00\xspace}
\newcommand{\chmuelelumiewkratio}{0.00\xspace}
\newcommand{\chmuelelumitop}{0.00\xspace}
\newcommand{\chmuelesystsnolumisminus}{0.13\xspace}
\newcommand{\chmuelesystsnolumisminuscorr}{0.12\xspace}
\newcommand{\chmuelesystsnolumisminusxsec}{3.97\xspace}
\newcommand{\chmuelesystsnolumisminuscorrxsec}{3.6\xspace}
\newcommand{\chmuelesystsnolumisplus}{0.08\xspace}
\newcommand{\chmuelesystsnolumispluscorr}{0.07\xspace}
\newcommand{\chmuelesystsnolumisplusxsec}{4.4\xspace}
\newcommand{\chmuelesystsnolumispluscorrxsec}{4.2\xspace}
\newcommand{\chmuelesystsnolumisratio}{0.10\xspace}
\newcommand{\chmuelesystsnolumisratiocorr}{0.10\xspace}
\newcommand{\chmuelesystsnolumisratioR}{0.00\xspace}
\newcommand{\chmuelesystsnolumisratioRcorr}{0.00\xspace}
\newcommand{\chmuelesystsnolumiewkmuminus}{0.00\xspace}
\newcommand{\chmuelesystsnolumiewkeleminus}{0.00\xspace}
\newcommand{\chmuelesystsnolumiewkratio}{0.00\xspace}
\newcommand{\chmuelesystsnolumitop}{0.00\xspace}
\newcommand{\chmusminus}{0.97\xspace}
\newcommand{\chmusminuscorr}{0.89\xspace}
\newcommand{\chmusminusxsec}{29.80\xspace}
\newcommand{\chmusminuscorrxsec}{27.3\xspace}
\newcommand{\chmusplus}{1.01\xspace}
\newcommand{\chmuspluscorr}{0.97\xspace}
\newcommand{\chmusplusxsec}{57.0\xspace}
\newcommand{\chmuspluscorrxsec}{54.68\xspace}
\newcommand{\chmusratio}{1.04\xspace}
\newcommand{\chmusratiocorr}{1.09\xspace}
\newcommand{\chmusratioR}{1.91\xspace}
\newcommand{\chmusratioRcorr}{2.00\xspace}
\newcommand{\chmuewkmuminus}{0.90\xspace}
\newcommand{\chmuewkeleminus}{0.00\xspace}
\newcommand{\chmuewkratio}{1.06\xspace}
\newcommand{\chmutop}{1.07\xspace}
\newcommand{\chmustatssminus}{0.06\xspace}
\newcommand{\chmustatssminuscorr}{0.05\xspace}
\newcommand{\chmustatssminusxsec}{1.79\xspace}
\newcommand{\chmustatssminuscorrxsec}{1.64\xspace}
\newcommand{\chmustatssplus}{0.04\xspace}
\newcommand{\chmustatsspluscorr}{0.03\xspace}
\newcommand{\chmustatssplusxsec}{2.00\xspace}
\newcommand{\chmustatsspluscorrxsec}{1.92\xspace}
\newcommand{\chmustatssratio}{0.06\xspace}
\newcommand{\chmustatssratiocorr}{0.07\xspace}
\newcommand{\chmustatssratioR}{0.12\xspace}
\newcommand{\chmustatssratioRcorr}{0.12\xspace}
\newcommand{\chmustatsewkmuminus}{0.12\xspace}
\newcommand{\chmustatsewkeleminus}{0.00\xspace}
\newcommand{\chmustatsewkratio}{0.05\xspace}
\newcommand{\chmustatstop}{0.06\xspace}
\newcommand{\chmusystssminus}{0.14\xspace}
\newcommand{\chmusystssminuscorr}{0.13\xspace}
\newcommand{\chmusystssminusxsec}{4.35\xspace}
\newcommand{\chmusystssminuscorrxsec}{3.99\xspace}
\newcommand{\chmusystssplus}{0.10\xspace}
\newcommand{\chmusystsspluscorr}{0.09\xspace}
\newcommand{\chmusystssplusxsec}{5.57\xspace}
\newcommand{\chmusystsspluscorrxsec}{5.35\xspace}
\newcommand{\chmusystssratio}{0.10\xspace}
\newcommand{\chmusystssratiocorr}{0.10\xspace}
\newcommand{\chmusystssratioR}{0.18\xspace}
\newcommand{\chmusystssratioRcorr}{0.18\xspace}
\newcommand{\chmusystsewkmuminus}{0.00\xspace}
\newcommand{\chmusystsewkeleminus}{0.00\xspace}
\newcommand{\chmusystsewkratio}{0.00\xspace}
\newcommand{\chmusyststop}{0.00\xspace}
\newcommand{\chmulumisminus}{0.03\xspace}
\newcommand{\chmulumisminuscorr}{0.02\xspace}
\newcommand{\chmulumisminusxsec}{0.77\xspace}
\newcommand{\chmulumisminuscorrxsec}{0.71\xspace}
\newcommand{\chmulumisplus}{0.03\xspace}
\newcommand{\chmulumispluscorr}{0.03\xspace}
\newcommand{\chmulumisplusxsec}{1.48\xspace}
\newcommand{\chmulumispluscorrxsec}{1.42\xspace}
\newcommand{\chmulumisratio}{0.00\xspace}
\newcommand{\chmulumisratiocorr}{0.00\xspace}
\newcommand{\chmulumisratioR}{0.00\xspace}
\newcommand{\chmulumisratioRcorr}{0.00\xspace}
\newcommand{\chmulumiewkmuminus}{0.00\xspace}
\newcommand{\chmulumiewkeleminus}{0.00\xspace}
\newcommand{\chmulumiewkratio}{0.00\xspace}
\newcommand{\chmulumitop}{0.00\xspace}
\newcommand{\chmusystsnolumisminus}{0.14\xspace}
\newcommand{\chmusystsnolumisminuscorr}{0.13\xspace}
\newcommand{\chmusystsnolumisminusxsec}{4.28\xspace}
\newcommand{\chmusystsnolumisminuscorrxsec}{3.92\xspace}
\newcommand{\chmusystsnolumisplus}{0.10\xspace}
\newcommand{\chmusystsnolumispluscorr}{0.09\xspace}
\newcommand{\chmusystsnolumisplusxsec}{5.37\xspace}
\newcommand{\chmusystsnolumispluscorrxsec}{5.15\xspace}
\newcommand{\chmusystsnolumisratio}{0.10\xspace}
\newcommand{\chmusystsnolumisratiocorr}{0.10\xspace}
\newcommand{\chmusystsnolumisratioR}{0.00\xspace}
\newcommand{\chmusystsnolumisratioRcorr}{0.00\xspace}
\newcommand{\chmusystsnolumiewkmuminus}{0.00\xspace}
\newcommand{\chmusystsnolumiewkeleminus}{0.00\xspace}
\newcommand{\chmusystsnolumiewkratio}{0.00\xspace}
\newcommand{\chmusystsnolumitop}{0.00\xspace}
\newcommand{\chelesminus}{1.01\xspace}
\newcommand{\chelesminuscorr}{0.91\xspace}
\newcommand{\chelesminusxsec}{30.87\xspace}
\newcommand{\chelesminuscorrxsec}{27.93\xspace}
\newcommand{\chelesplus}{0.99\xspace}
\newcommand{\chelespluscorr}{0.92\xspace}
\newcommand{\chelesplusxsec}{55.91\xspace}
\newcommand{\chelespluscorrxsec}{51.86\xspace}
\newcommand{\chelesratio}{0.99\xspace}
\newcommand{\chelesratiocorr}{1.01\xspace}
\newcommand{\chelesratioR}{1.81\xspace}
\newcommand{\chelesratioRcorr}{1.86\xspace}
\newcommand{\cheleewkmuminus}{0.00\xspace}
\newcommand{\cheleewkeleminus}{1.10\xspace}
\newcommand{\cheleewkratio}{0.98\xspace}
\newcommand{\cheletop}{1.02\xspace}
\newcommand{\chelestatssminus}{0.08\xspace}
\newcommand{\chelestatssminuscorr}{0.07\xspace}
\newcommand{\chelestatssminusxsec}{2.43\xspace}
\newcommand{\chelestatssminuscorrxsec}{2.20\xspace}
\newcommand{\chelestatssplus}{0.05\xspace}
\newcommand{\chelestatsspluscorr}{0.04\xspace}
\newcommand{\chelestatssplusxsec}{2.58\xspace}
\newcommand{\chelestatsspluscorrxsec}{2.40\xspace}
\newcommand{\chelestatssratio}{0.09\xspace}
\newcommand{\chelestatssratiocorr}{0.09\xspace}
\newcommand{\chelestatssratioR}{0.16\xspace}
\newcommand{\chelestatssratioRcorr}{0.16\xspace}
\newcommand{\chelestatsewkmuminus}{0.00\xspace}
\newcommand{\chelestatsewkeleminus}{0.16\xspace}
\newcommand{\chelestatsewkratio}{0.07\xspace}
\newcommand{\chelestatstop}{0.05\xspace}
\newcommand{\chelesystssminus}{0.15\xspace}
\newcommand{\chelesystssminuscorr}{0.13\xspace}
\newcommand{\chelesystssminusxsec}{4.56\xspace}
\newcommand{\chelesystssminuscorrxsec}{4.12\xspace}
\newcommand{\chelesystssplus}{0.09\xspace}
\newcommand{\chelesystsspluscorr}{0.08\xspace}
\newcommand{\chelesystssplusxsec}{5.13\xspace}
\newcommand{\chelesystsspluscorrxsec}{4.76\xspace}
\newcommand{\chelesystssratio}{0.13\xspace}
\newcommand{\chelesystssratiocorr}{0.13\xspace}
\newcommand{\chelesystssratioR}{0.23\xspace}
\newcommand{\chelesystssratioRcorr}{0.24\xspace}
\newcommand{\chelesystsewkmuminus}{0.00\xspace}
\newcommand{\chelesystsewkeleminus}{0.00\xspace}
\newcommand{\chelesystsewkratio}{0.00\xspace}
\newcommand{\chelesyststop}{0.00\xspace}
\newcommand{\chelelumisminus}{0.03\xspace}
\newcommand{\chelelumisminuscorr}{0.02\xspace}
\newcommand{\chelelumisminusxsec}{0.80\xspace}
\newcommand{\chelelumisminuscorrxsec}{0.73\xspace}
\newcommand{\chelelumisplus}{0.03\xspace}
\newcommand{\chelelumispluscorr}{0.02\xspace}
\newcommand{\chelelumisplusxsec}{1.45\xspace}
\newcommand{\chelelumispluscorrxsec}{1.35\xspace}
\newcommand{\chelelumisratio}{0.00\xspace}
\newcommand{\chelelumisratiocorr}{0.00\xspace}
\newcommand{\chelelumisratioR}{0.00\xspace}
\newcommand{\chelelumisratioRcorr}{0.00\xspace}
\newcommand{\chelelumiewkmuminus}{0.00\xspace}
\newcommand{\chelelumiewkeleminus}{0.00\xspace}
\newcommand{\chelelumiewkratio}{0.00\xspace}
\newcommand{\chelelumitop}{0.00\xspace}
\newcommand{\chelesystsnolumisminus}{0.15\xspace}
\newcommand{\chelesystsnolumisminuscorr}{0.13\xspace}
\newcommand{\chelesystsnolumisminusxsec}{4.48\xspace}
\newcommand{\chelesystsnolumisminuscorrxsec}{4.06\xspace}
\newcommand{\chelesystsnolumisplus}{0.09\xspace}
\newcommand{\chelesystsnolumispluscorr}{0.08\xspace}
\newcommand{\chelesystsnolumisplusxsec}{4.92\xspace}
\newcommand{\chelesystsnolumispluscorrxsec}{4.56\xspace}
\newcommand{\chelesystsnolumisratio}{0.13\xspace}
\newcommand{\chelesystsnolumisratiocorr}{0.13\xspace}
\newcommand{\chelesystsnolumisratioR}{0.00\xspace}
\newcommand{\chelesystsnolumisratioRcorr}{0.00\xspace}
\newcommand{\chelesystsnolumiewkmuminus}{0.00\xspace}
\newcommand{\chelesystsnolumiewkeleminus}{0.00\xspace}
\newcommand{\chelesystsnolumiewkratio}{0.00\xspace}
\newcommand{\chelesystsnolumitop}{0.00\xspace}
\newcommand{\VtbValue}{0.979\xspace}
\newcommand{\VtbValueUnc}{0.045\xspace}

\newcommand{\VtbValueComb}{0.998\xspace}
\newcommand{\VtbValueUncComb}{0.038\xspace}

\newcommand{\RatioSevenEight}{1.24\xspace}
\newcommand{\RatioSevenEightUnc}{0.14\xspace}
\newcommand{\RatioSevenEightUncSyst}{0.12\xspace}
\newcommand{\RatioSevenEightUncStat}{0.08\xspace}
\cmsNoteHeader{TOP-12-038} 
\title{Measurement of the $t$-channel single-top-quark production cross section and of the $\abs{\vtb}$ CKM matrix element in pp collisions at $\sqrt{s}= 8\TeV$}

\date{\today}

\abstract{
Measurements are presented of the $t$-channel single-top-quark production cross section in proton-proton collisions at $\sqrt{s}=8$\TeV. The results are based on a data sample corresponding to an integrated luminosity of 19.7\fbinv recorded with the CMS detector at the LHC. The cross section is measured inclusively, as well as separately for top (\tq) and antitop (\tqbar), in final states with a muon or an electron. The measured inclusive $t$-channel cross section is $\sigma_{\tch} = 83.6 \pm 2.3\stat \pm 7.4\syst\unit{pb}$. The single \tq and \tqbar cross sections are measured to be $\sigma_{\tch}(\cPqt) = 53.8 \pm 1.5\stat\pm 4.4\syst\unit{pb}$ and $\sigma_{\tch}(\cPaqt) = 27.6 \pm 1.3\stat\pm 3.7\syst\unit{pb}$, respectively.
The measured ratio of cross sections is $R_{\tch}=\sigma_{\tch}(\cPqt)/\sigma_{\tch}(\cPaqt)=1.95 \pm 0.10\stat\pm 0.19\syst$, in agreement with the standard model prediction. The modulus of the Cabibbo--Kobayashi--Maskawa matrix element $\Vtb$ is extracted and, in combination with a previous CMS result at $\sqrt{s} = 7\TeV$, a value $\absvtb = 0.998\pm0.038\exper\pm 0.016\thy$ is obtained.}

\hypersetup{%
pdfauthor={CMS Collaboration},%
pdftitle={Measurement of the t-channel single-top-quark production cross section and of the |Vtb| CKM matrix element in pp collisions at sqrt(s) = 8 TeV},%
pdfsubject={CMS},%
pdfkeywords={CMS, top quark, single top}}

\maketitle 

\section{Introduction}
\label{sec:intro}
\label{sec:introduction}
\label{sec:Introduction}
\begin{figure}[ht]
  \centering
    \includegraphics[width=0.36\textwidth]{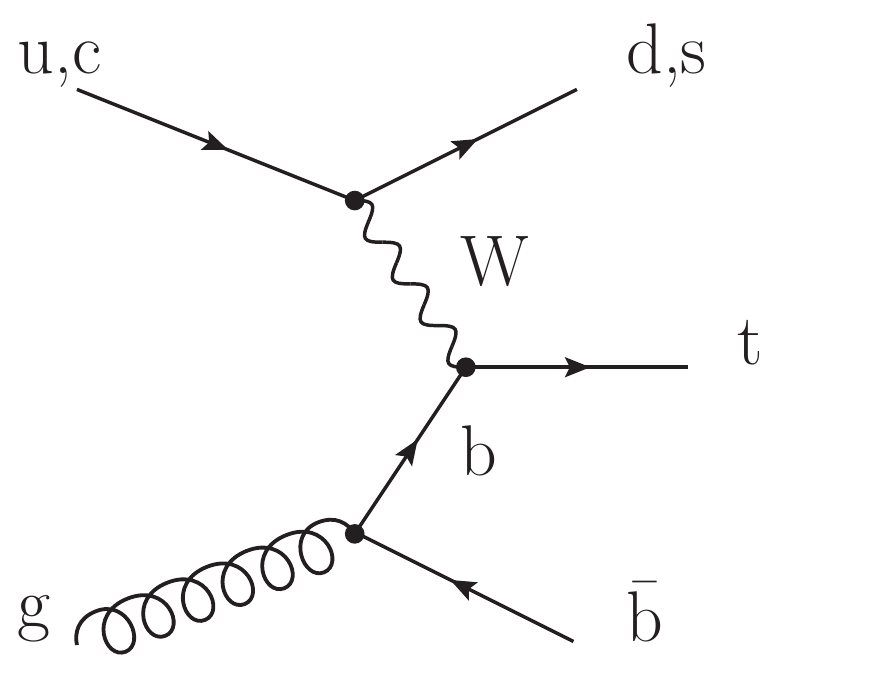}
    \includegraphics[width=0.36\textwidth]{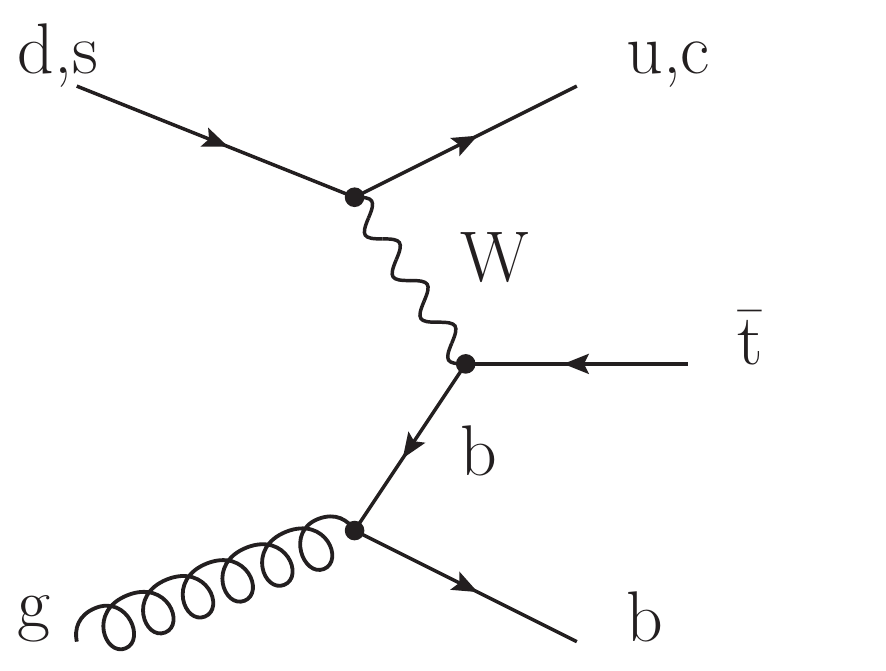}
    \caption{\label{fig:FG} Leading-order Feynman diagrams for (left) single \tq and (right) \tqbar production in the $t$-channel. }
\end{figure}

In the standard model (SM), the production of single top quarks (\tq or its antiparticle \tqbar) in proton-proton (pp) collisions proceeds through the charged-current electroweak interaction.
At leading order (LO), three different mechanisms can be distinguished, namely the $t$-channel, the $s$-channel and the associated production of a single top quark and a W boson (tW)~\cite{firstharris,firstkidonakis,firsttramontano}.
In this work, measurements are presented of $t$-channel production. LO diagrams contributing to $t$-channel
single \tq and \tqbar production are presented in figure~\ref{fig:FG}.
Processes involving single top quarks provide direct probes of electroweak interactions, and thereby important tests of the SM predictions
as well as excellent opportunities for searching for new physics. Since a Wtb vertex, where W and b denote the W boson and the b quark respectively, is involved in all SM single-top-quark production mechanisms, the modulus of the Cabibbo--Kobayashi--Maskawa (CKM) matrix element $\abs{\Vtb}$ can be determined from their measured cross sections.
Depending on whether the b quarks are considered  part of the proton or not, single-top-quark production can be studied in the 5- or 4-flavour schemes~\cite{fourfiveflavorschemes}, respectively.
In the 4-flavour (4F) scheme, the b quarks are generated in the hard scattering from gluon splitting. In the 5-flavour (5F) scheme, the b quarks are considered as constituents of the proton.
An additional feature of the $t$- and $s$-channels, specific to pp collisions, is the difference between production cross sections of
single \tq and \tqbar that results from a difference in parton distribution functions (PDF) of incident up and down quarks involved in the hard scattering.
The ratio of \tq over \tqbar production cross sections in the $t$-channel ($R_{\tch}$) is therefore sensitive to the PDF of the up- and down-type quarks in the proton. The ratio $R_{\tch}$ is also directly sensitive to physics beyond the SM manifesting as anomalous couplings in the Wtb vertex~\cite{twbBSM}, or to possible contributions from flavour-changing neutral current processes~\cite{fcncBSM}.

For pp collisions at a centre-of-mass energy $\sqrt{s}= 8\TeV$, the predicted theoretical cross section for SM $t$-channel single-top-quark production is

\begin{equation}
\sigma_{\tch}^\text{theo.} = 87.2^{+2.8}_{-1.0}\,\text{(scale)}^{+2.0}_{-2.2}\,\text{(PDF)}\unit{pb},
\label{eq:inclusivexsectheor}
\end{equation}

as obtained in quantum chromodynamics (QCD) at approximate next-to-next-to-leading order (NNLO) including resummation of the soft-gluon emission with the next-to-next-to-leading-logarithms (NNLL) calculation~\cite{Kidonakis:2012db}.
The PDF set MSTW08NNLO~\cite{MSTW2008NLO} is used in the 5F scheme.
For this calculation the top-quark mass $m_{\cPqt}$ is set to 173\GeV, and the factorisation and renormalisation scales are set both to $m_{\cPqt}$. The uncertainty receives contributions from the PDF uncertainty and the missing higher-order corrections, estimated by varying the factorisation and renormalisation scales by a multiplicative factor of 0.5 or 2.0.
The same calculations predict the following production cross sections for single \tq and \tqbar, separately:
\begin{equation}\begin{aligned}
\sigma_{\tch}^\text{theo.}(\cPqt) & = 56.4^{+2.1}_{-0.3}\,\text{(scale)}\pm1.1\,\text{(PDF)}\unit{pb}, \\
\sigma_{\tch}^\text{theo.}(\cPaqt) & = 30.7\pm0.7\,\text{(scale)}^{+0.9}_{-1.1}\,\text{(PDF)}\unit{pb}.
\end{aligned}\end{equation}

Single-top-quark events were observed for the first time in proton-antiproton collisions at $\sqrt{s}=1.96\TeV$
at the Tevatron~\cite{CDF-singletop,D0-singletop}.
At the Large Hadron Collider (LHC) both ATLAS and CMS collaborations observed production of single-top-quark events
in the $t$-channel in pp collisions at $\sqrt{s}=7\TeV$~\cite{Chatrchyan:2012vp,Chatrchyan:2011vp,ATLAS-singletop}.
Single-top-quark tW production has been recently observed at $\sqrt{s}=8\TeV$ by the Compact Muon Solenoid (CMS)
collaboration~\cite{CMS-tW-Obs}, while observation of $s$-channel production was reported at the Tevatron~\cite{D0-s}.

The measurement performed by CMS of inclusive single-top-quark production cross section in the $t$-channel
at $\sqrt{s}=8\TeV$, as well as separate measurements of single \tq and \tqbar production cross sections are presented.
Signal events are characterised by products of top-quark decay that are accompanied by a light quark emitted at high rapidity and a soft b quark. 
Events are selected if a muon or electron consistent with originating from a top-quark decay chain is present in the final state.
The signal yield is extracted from a maximum likelihood fit to the distribution of the absolute value of the pseudorapidity ($\eta$) of the jet ($j'$) originating from the light quark, $\absetalj$. The expected distributions of $\absetalj$ are determined from data for the relevant backgrounds.
Two independent fit procedures are implemented to extract the total $t$-channel production cross section
and \tq and \tqbar production cross sections separately.
The ratio of $t$-channel production cross sections at $\sqrt{s}=8\TeV$ and 7\TeV, $R_{8/7}$, can provide complementary information on the PDF with respect to the ratio of \tq over \tqbar, and can be compared to the prediction obtained using the cross sections in ref.~\cite{Kidonakis:2012db}, which is:
\begin{eqnarray}
R_{8/7}^\text{theo.}  = 1.32^{+0.06}_{-0.02}\,\text{(scale)}^{+0.04}_{-0.05}\,\text{(PDF)}.
\end{eqnarray}


\section{The CMS detector}
\label{sec:cms}
\label{sec:CMS}
The CMS apparatus features a 6\unit{m} diameter superconducting solenoid
that provides a magnetic field of 3.8\unit{T} and allows for the relatively compact design of the detector.
The inner bore of the solenoid hosts a tracking system, composed of silicon pixel and silicon strip detectors, that allows for reconstruction of charged-particle
tracks bending in the internal magnetic field. A lead tungstate crystal electromagnetic
calorimeter and a brass/scintillator hadron calorimeter surround the tracker volume. Outside the
solenoid, gas-ionisation detectors, \ie resistive plate chambers, drift tubes and cathode strip chambers, are interleaved
with the steel flux-return yoke of CMS and form the muon
system. A quartz-fibre and steel absorber Cherenkov calorimeter,
located outside the muon system close to the beam pipe, extends the calorimetric system
angular acceptance in the region along the beam axis. A more detailed description of the CMS detector can be found in ref.~\cite{Chatrchyan:2008zzk}.

The CMS experiment uses a right-handed coordinate system centred on the nominal interaction point, with the $z$-axis
parallel to the anticlockwise-beam direction, the $x$ axis lying in the plane of the LHC ring and pointing to its centre,
and the $y$ axis pointing upwards to the surface. The pseudorapidity $\eta$ is defined as $-\ln{ [ \tan{ ( \theta/2 ) } ] }$,
where $\theta$ is the azimuthal angle with respect to the $z$ axis.

\section{Data and simulated samples}
\label{sec:datasets}
The measurement is performed on a data sample collected during 2012
at $\sqrt{s}=8\TeV$, selected with triggers requiring one muon ($\mu$) or one electron (e),
and corresponding to an integrated luminosity of 19.7\fbinv.

The simulated $t$-channel events are generated with
\POWHEG~1.0~\cite{Re:2010bp,Alioli:2010xd,Alioli:2009je,Frixione:2007vw} interfaced to {\PYTHIA~6.4 ~\cite{Sjostrand:2006za} for parton shower evolution and hadronisation.
Other single-top-quark processes, i.e. the $s$-channel and the tW, are considered as backgrounds for this measurement and simulated
with the same Monte Carlo (MC) generators. Top quark pair production, single vector boson production
associated with jets (\vjets), and double vector boson (diboson) production are amongst the backgrounds taken into consideration and have been
simulated with {\MADGRAPH~5.148}~\cite{madgraph} interfaced to \PYTHIA for parton showering.
The \PYTHIA generator is used to simulate \QCD samples enriched with isolated muons or electrons.
The value of the top-quark mass used in all simulated samples is $m_{\cPqt}=$172.5\GeV.
All samples are generated using the CTEQ6.6M~\cite{CTEQ66} PDF set. The factorisation and renormalisation scales are both set
to $m_{\cPqt}$ for the single-top-quark samples, while a dynamic scale is used for the other samples.
The  production cross section used to scale the simulation of single-top-quark tW and $s$-channel processes is taken from ref.~\cite{Kidonakis:2012db},
while the \ttbar production cross section is taken from ref.~\cite{ttbarxsec}.

\section{Event selection and reconstruction}
\label{sec:selection}
The signal events are defined by the decay of $\ttoleptoncascade$, where $\ell=\mu,\Pe$. The $\ttotaucascade$ contribute to the signal
when a $\tau$ decays leptonically. The resulting final state includes a muon or electron, and escaping neutrinos ($\nu$) that cause an imbalance in the momentum measured in the transverse plane. A bottom (or b) jet that stems from the hadronisation of the b quark from the top-quark decay accompanies the leptons. An additional jet originates from the light-flavoured quark recoiling against the top quark. The splitting of the gluon from the initial state produces a second b quark that recoils against the top quark, as shown in figure~\ref{fig:FG}. The b jet from gluon splitting has generally a softer transverse momentum ($\pt$) spectrum and a broader $\abs{\eta}$ distribution
compared to the one produced in top-quark decay, thus the acceptance for events with two b jets reconstructed in the final
state is relatively small. In fact we can anticipate that using the selection described in this section, the number of signal events with two b jets reconstructed in the detector is one order of magnitude smaller than the number of events with just one  b jet.

Events are selected online by the high-level trigger system requiring the presence of either one isolated muon with
 $\pt>24\GeV$ and pseudorapidity $\abs{\eta}<2.1$, or one isolated electron with $\pt>27\GeV$ and $\abs{\eta}<2.5$.
The event is required to have at least one primary vertex reconstructed from at least four tracks, with a distance
from the nominal beam-interaction point of less than 24\unit{cm} along the $z$ axis and less than 2\unit{cm} in the transverse plane.
When more than one primary vertex is found, the one with the largest $\sum \pt^2$ is used as leading vertex.
All particles are reconstructed and identified with the CMS particle-flow (PF) algorithm~\cite{CMS-PAS-PFT-09-001,CMS-PAS-PFT-10-002}.
Events with exactly one good muon or electron candidate are accepted for analysis.
Good muon candidates must have $\pt >26\GeV$ and $\abs{\eta} < 2.1$, while
electron candidates must have $\pt >30\GeV$ and $\abs{\eta} < 2.5$, excluding the barrel-endcap transition region $1.44 < \abs{\eta} < 1.57$ because the reconstruction of an electron in this region is not optimal.
The $\pt$ requirements on the leptons ensure that selected muons and electrons are in the plateau region of the respective trigger turn-on curves.

Muon isolation is ensured by applying requirements on the variable $\PFRelIso$, defined as the sum of the transverse energies deposited
by stable charged hadrons, photons, and neutral hadrons in a cone of size $\Delta R = \sqrt{\smash[b]{(\Delta\eta)^2+(\Delta\phi)^2}} = 0.4$, (where $\phi$ is the polar angle in radians) corrected by the average contribution of neutral particles from overlapping pp interactions (pileup), and divided by the muon $\pt$. Muons are required to have $\PFRelIso<0.12$.
Electron isolation criteria are based on a variable defined analogously to the muons, with an isolation cone of size $\Delta R = 0.3$. The isolation requirement for electrons is $\PFRelIso <$ 0.1.
Events are rejected if an additional muon (electron) candidate is present, passing looser selection requirements of
$\pt > 10\,(20)\GeV$, $\abs{\eta} < 2.5$ (including the barrel-endcap transition region for electrons), and $\PFrelIso <0.2\,(0.15)$.

The missing transverse momentum vector $\mpt$ is defined as the negative vector sum of the transverse momenta of
all reconstructed particles. The missing transverse energy $\ETslash$
is defined as the magnitude of $\mpt$.
The transverse mass $\mTW$ for events with a muon is calculated as
\begin{equation}
\mTW = \sqrt{\left(\pt^{\mu} + \ETslash\right)^2 - \left( p_x^{\mu} + \mpx \right)^2 - \left( p_y^{\mu} + \mpy \right)^2},
\end{equation}
 where $p_x^{\mu}$ and $p_y^{\mu}$ are the component of the muon momentum along the $x$ and $y$ axes, and $\mpx$ and $\mpy$ are the components of $\mpt$ along the $x$ and $y$ axes.
In order to reduce the \QCD background, a requirement of $\mTW>50\GeV$ is applied for the muon decay channel, while
a requirement of $\ETslash > 45$\GeV is applied instead for the electron channel.
Control region studies, described in section~\ref{sec:qcd}, show that the procedure
for the \QCD extraction in the electron channel yields a considerably smaller uncertainty when applying the
requirement on $\ETslash$ rather than on $\mTW$.

Jets are defined by clustering reconstructed particles with the anti-\kt algorithm~\cite{Cacciari:2008gp} with a distance parameter of 0.5. Charged particles are excluded if they have a distance with respect to any primary vertex along the $z$ axis smaller than that with respect to the leading vertex.
The average energy density in $\eta$-$\phi$ space of neutral particles
not clustered into jets is used to extrapolate the energy due to pileup
interactions in the jet cone. The jet energy is corrected accordingly.
Further jet energy corrections are derived from the study of dijet events and photon+jets events (see ref.~\cite{Chatrchyan:2011ds}).
Jets are required to be within $\abs{\eta}<4.5$ and to have a transverse energy $\et >40\GeV$.
In order to identify b-quark-induced jets, a b-tagging algorithm is used
exploiting the 3D impact parameter of the tracks in the jet to define a ``b-discriminator''~\cite{CMS-PAS-BTV-11-004}.
An optimised threshold is chosen on this variable with probability to misidentify jets coming from the hadronisation of light quarks (u, d, s) or gluons of 0.3\% and an efficiency of selecting jets coming from b quarks of 46\%, determined from simulation.
Jets passing the chosen threshold are considered as ``b-tagged''.  The majority of the background events surviving the final selection contain an actual b jet, the main exception being W+c-jet events. This algorithm is found to have good discriminating power with respect to this particular background.

Events are divided into categories according to the number of jets and b-tagged jets using the wording ``$n$-jet $m$-tag '', referring to events with $n$ jets,
$m$ of which are b-tagged.
Once the event is been assigned to a category, a further selection based on the jet shape is performed to reduce the
contamination due to jets coming from pileup interactions:
the distance in the $\eta$-$\phi$ plane between the momenta of the particles constituting the jet and the jet axis is evaluated and
its root mean square (\textsc{rms}) over all the jet constituents is required to be smaller than 0.025.
This requirement is applied on the jets that are not classified as b-tagged, and the event is rejected if either of those jets does not satisfy it.
This requirement allows us to discriminate jets coming from u and d quarks with
respect to jets coming from gluons or b quarks, which present a broader jet profile.
This criteria 
is particularly useful in the forward region of the detector where other quality criteria making use of the tracking system cannot be applied.
A top-quark candidate is reconstructed in the 2-jet 1-tag, in the 3-jet 1-tag  and in the 3-jet 2-tag samples from a lepton, \ETslash and one
b jet combination with the algorithm described in ref.~\cite{Chatrchyan:2012vp}. The b jet with the highest value of the b-discriminator is used for top-quark reconstruction in the 3-jet 2-tag sample.
The mass of such a candidate ``$\mt$'' is used to define a signal region (SR) and a sideband region (SB) in each of those samples, selecting events respectively inside and outside the reconstructed top-quark mass window of $130 < \mt <220$\GeV.
The variable $\etalj$ in the 2-jet 0-tag sample is defined taking the pseudorapidity of each of the two jets, and two entries per event are present.
In the 3-jet 1-tag sample $\etalj$ is defined as the pseudorapidity of the jet with the smallest b-discriminator value. In the 2-jet 1-tag and in the 3-jet 2-tag samples it is defined as the pseudorapidity of the non-b-tagged jet.
The category enriched with $t$-channel signal is the one with 2 jets and 1 tag.
The final procedure to isolate the signal from background uses the absolute value $\absetalj$.
The pseudorapidity distribution of the outgoing jet $j'$ is typical of the $t$-channel processes where a light parton recoils against a much more massive particle like the top quark.
Signal events populate forward regions in the $\absetalj$ spectrum that correspond to the tails of the $\absetalj$ distribution for SM processes.

The total event yields in the signal and sideband regions of the 2-jet 1-tag sample for muons and electrons are reported in table~\ref{tab:yield_tot}.
The event yields in the signal region for positively and negatively charged muons and electrons separately are reported in table~\ref{tab:yield_ch}.

 \begin{table}[ht]
\topcaption{Event yield for the main processes in the 2-jet 1-tag signal region (SR) and sideband region (SB), for the muon and electron decay channels. Expected yields are taken from simulation and their uncertainties are due to the finite size of the MC sample with the exception of \QCD yield (see section~\ref{sec:qcd}), and $\wzjets$ yield (see section~\ref{sec:whfextraction}), whose yields and uncertainties are taken as the statistical component of the uncertainty in the estimation from data.}
 \centering
 \begin{tabular}{ |c|c|c||c|c| }
\hline
Process & \multicolumn{2}{ c ||} {Muon} & \multicolumn{2}{ c |}{Electron}  \\
  & \multicolumn{1}{ c } { SR } & \multicolumn{1}{ c ||} { SB } &  \multicolumn{1}{ c } { SR } & \multicolumn{1}{ c |} { SB }  \\
\hline
\ttbar & 17214 $\pm$ 49 \,  & 8238 $\pm$ 35 & 11162 $\pm$ 38 \, & 8036 $\pm$ 33 \\
\wzjets & 10760 $\pm$ 104 & 9442 $\pm$ 97 & 4821 $\pm$ 69 & 6512 $\pm$ 81 \\
\QCD & 765 $\pm$ 5 & 271 $\pm$ 4 & 1050 $\pm$ 6 \, & 1350 $\pm$ 6 \, \\
Diboson & 179 $\pm$ 4 & 161 $\pm$ 4 & \, 95 $\pm$ 3 & 134 $\pm$ 3 \\
tW & 1914 $\pm$ 28 &  \, 969 $\pm$ 20 & 1060 $\pm$ 28 & \, 858 $\pm$ 18 \\
$s$-channel & 343 $\pm$ 1 & 118 $\pm$ 1 & 180 $\pm$ 1 & \, 96 $\pm$ 1 \\
\hline
$t$-channel & 6792 $\pm$ 25 & 944 $\pm$ 9 & 3616 $\pm$ 17 & 753 $\pm$ 8 \\
\hline
Total expected & 37967 $\pm$ 121 &  20143 $\pm$ 106 & 21984 $\pm$  85 \, & 17740 $\pm$ 90 \, \\
\hline
Data & 38202  & 20237 & 22597  & 17700 \\
\hline
 \end{tabular}
\label{tab:yield_tot}
 \end{table}

 \begin{table}[ht]
\topcaption{Event yield for the main processes in the 2-jet 1-tag signal region, for events with positively and negatively charged muons and electrons. Expected yields are taken from simulation and their uncertainties are due to the finite size of the MC sample with the exception of \QCD yield (see section~\ref{sec:qcd}), and $\wzjets$ yield (see section~\ref{sec:whfextraction}), whose yields and uncertainties are taken as the statistical component of the uncertainty in the estimation from data.}
 \centering
 \begin{tabular}{ |c|c|c||c|c| }
\hline
Process & \multicolumn{2}{ c ||} {Muon} & \multicolumn{2}{ c |}{Electron}  \\
  & \multicolumn{1}{ c } {+} & \multicolumn{1}{ c ||} {$-$} &  \multicolumn{1}{ c } {+} & \multicolumn{1}{ c |} {$-$}  \\
\hline
\ttbar & 8620 $\pm$ 35 & 8594 $\pm$ 35  & 5574 $\pm$ 27 & 5588 $\pm$ 27 \\
\wzjets & 5581 $\pm$ 75 & 4989 $\pm$ 71 & 2618 $\pm$ 52 & 2121 $\pm$ 46 \\
\QCD & 361 $\pm$ 1 & 366 $\pm$ 1 & 697 $\pm$ 2 & 679 $\pm$ 2 \\
Diboson & 106 $\pm$ 3 & \, 73 $\pm$ 2  & \, 58 $\pm$ 2 &\, 39 $\pm$ 2 \\
tW &\, 964 $\pm$ 20 & \, 951 $\pm$ 20 & \, 535 $\pm$ 14 &\, 525 $\pm$ 14 \\
$s$-channel & 225 $\pm$ 1 & 118 $\pm$ 1 & 118 $\pm$ 1 & \, 62 $\pm$ 1 \\
\hline
$t$-channel & 4325 $\pm$ 19 & 2467 $\pm$ 16  & 2320 $\pm$ 13 & 1295 $\pm$ 11 \\
\hline
Total expected & 20181 $\pm$ 87 \, & 17557 $\pm$ 83 \, & 11920 $\pm$ 61 \,& 10310 $\pm$ 56 \,\\
\hline
Data & 20514 & 17688  & 12035 & 10562 \\
\hline
 \end{tabular}
\label{tab:yield_ch}
 \end{table}

\section{Background estimation and control samples}
\label{sec:bkg}
\label{sec:backgrounds}
\label{sec:control}
The physics processes that constitute the main backgrounds to single-top-quark production in the $t$-channel are $\ttbar$, \wjets, and \QCD production.
Control samples are defined for each of these contributions in order to check that the variables used in the analysis are reproduced correctly in the simulations.
For the main backgrounds the most important distributions, together with constraints on their production rates, are derived from data making use of these control samples.

\subsection{QCD multijet background}
\label{sec:qcd}
The vast majority of \QCD events are successfully rejected applying the selection described in section~\ref{sec:selection}.
The selected multijet events are thus found to be rare occurrences in the respective distributions,
for instance populating the tails of the typical multijet lepton-$\pt$ spectra.
The modelling uncertainties on the simulation have greater impact in those regions.
We thus estimate the \QCD contribution in our signal and sideband regions directly
from data, in the 2-jet 1-tag category as well as in the other control samples.
The measurement is performed with a fit to the distribution of the transverse mass in the muon decay channel, and to the distribution of the missing transverse energy in the electron decay channel.
A maximum-likelihood fit to the distribution either of \mTW in the muon case, or \ETslash in the electron case is
performed. The data are parametrised as: $F_\ell(x)= a_\ell\cdot S_\ell(x)+(1-a_\ell)\cdot B_\ell(x)$ for $\ell = \mu$, \Pe.
The variable $x$ is \mTW for the muon decay channel
and \ETslash for the electron decay channel, while $S_\ell(x)$ and $B_\ell(x)$ are the expected distributions for the sum of all processes with a W boson and \QCD events, respectively. The distribution $S_\ell(x)$ is derived from simulation and it includes the contribution from the signal.
The distribution $B_\ell(x)$ is obtained from a \QCD enriched data sample defined by taking muons and electrons with the same criteria as defined in section~\ref{sec:selection}, but with reversed isolation requirements for both leptons, selecting muons or electrons with $\PFrelIso> 0.2$ or 0.15 respectively.
The data samples defined in this way contain a fraction
of events originating from \QCD processes of 98\%
in the case of the muon decay channel and of more that 99\% for the electron decay channel.
The residual contribution from other non-\QCD processes is
subtracted from these samples using the expectation from simulation.
The fit procedure is repeated using different \QCD models, obtained by either varying the isolation requirement that defines the control
region or using the simulation for the \QCD distribution.
The kinematic bias on the multijet \mTW(\ETslash) distributions due to the extraction from the control sample is
covered by the systematic uncertainty defined this way.
\subsection{ Top quark pair background}
\label{sec:ttbarextraction}

The \ttbar process dominates in events with larger jet and b-tag multiplicity than the 2-jet 1-tag sample used for signal extraction.
Two control samples enriched in $\ttbar$ are thus defined, labelled 3-jet 1-tag and 3-jet 2-tag.
The distribution of $\absetalj$ in the 3-jet 1-tag and in the 3-jet 2-tag samples is shown in figure~\ref{fig:ttbar_eta}.
Good agreement between data and simulation in the two control samples is displayed, giving confidence in the simulation of the kinematic properties of the $\ttbar$ background.
The lepton charge in the 3-jet 1-tag and 3-jet 2-tag samples is shown in figure~\ref{fig:ttbar_charge}. The corresponding charge ratio
in the two samples is shown in figure~\ref{fig:ttbar_charge_ratio}, and is close to unity as expected for $\ttbar$ enriched samples.

\begin{figure}[!ht]
\centering
\includegraphics[width=\cmsFigWidth]{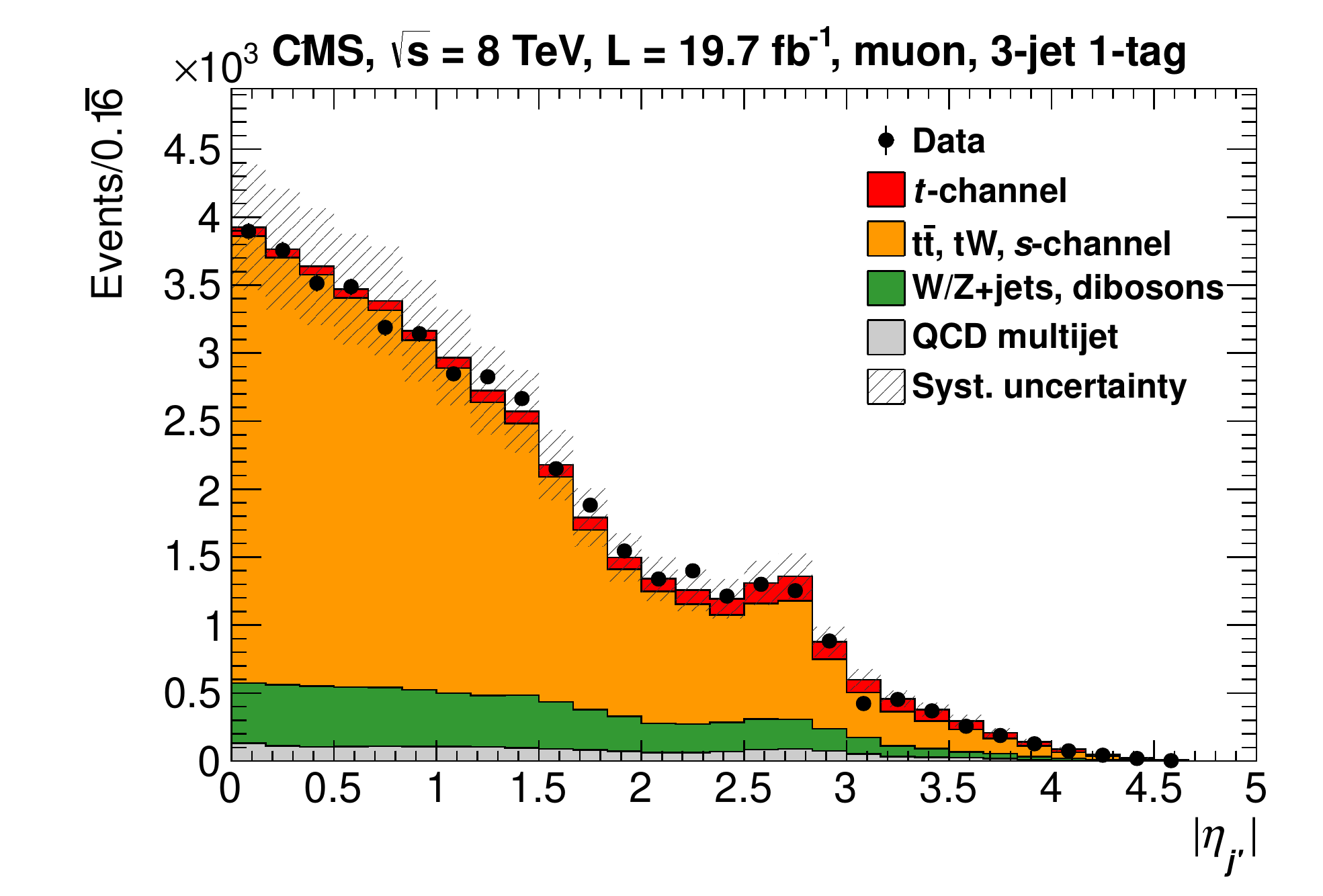}
\includegraphics[width=\cmsFigWidth]{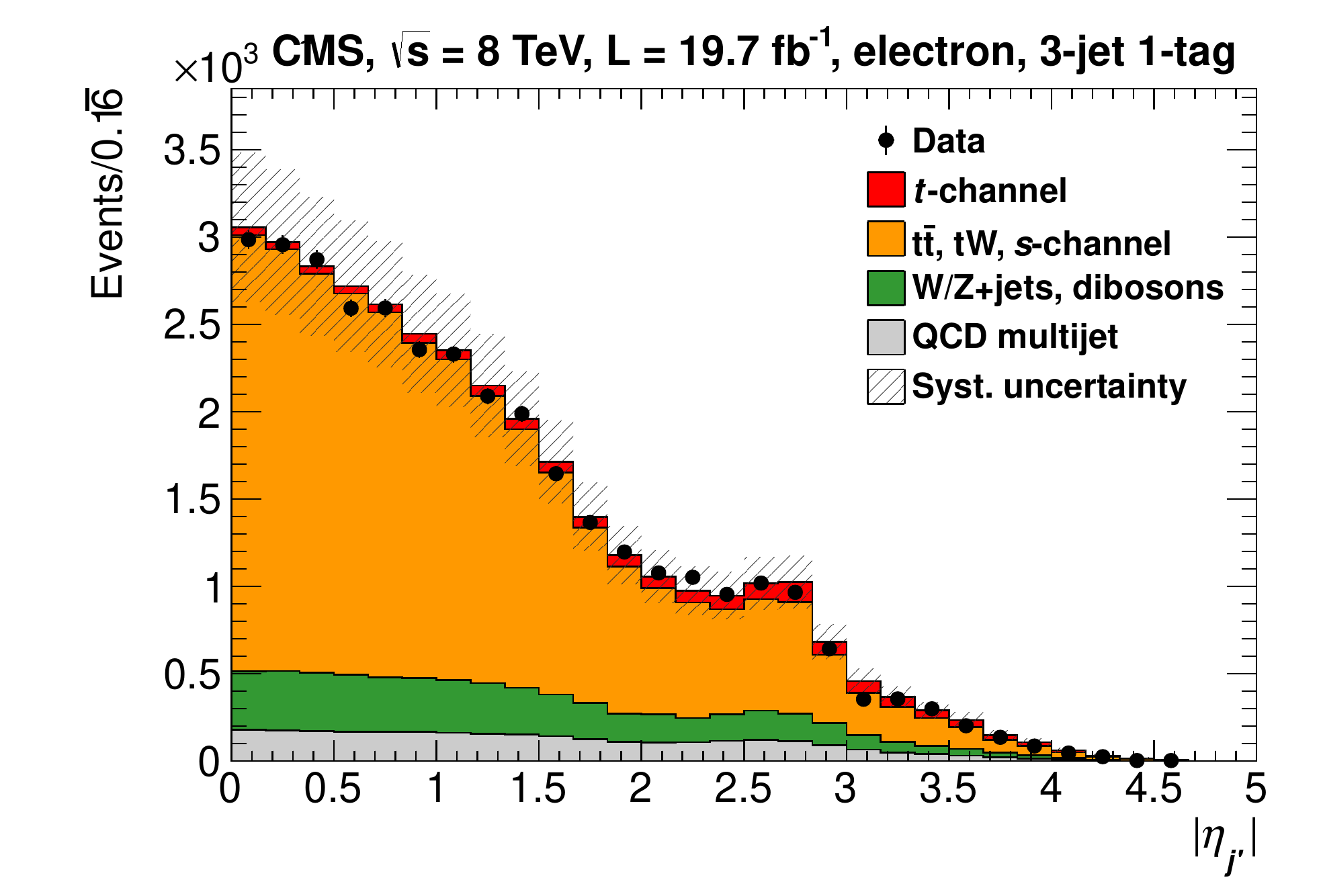}\\
\includegraphics[width=\cmsFigWidth]{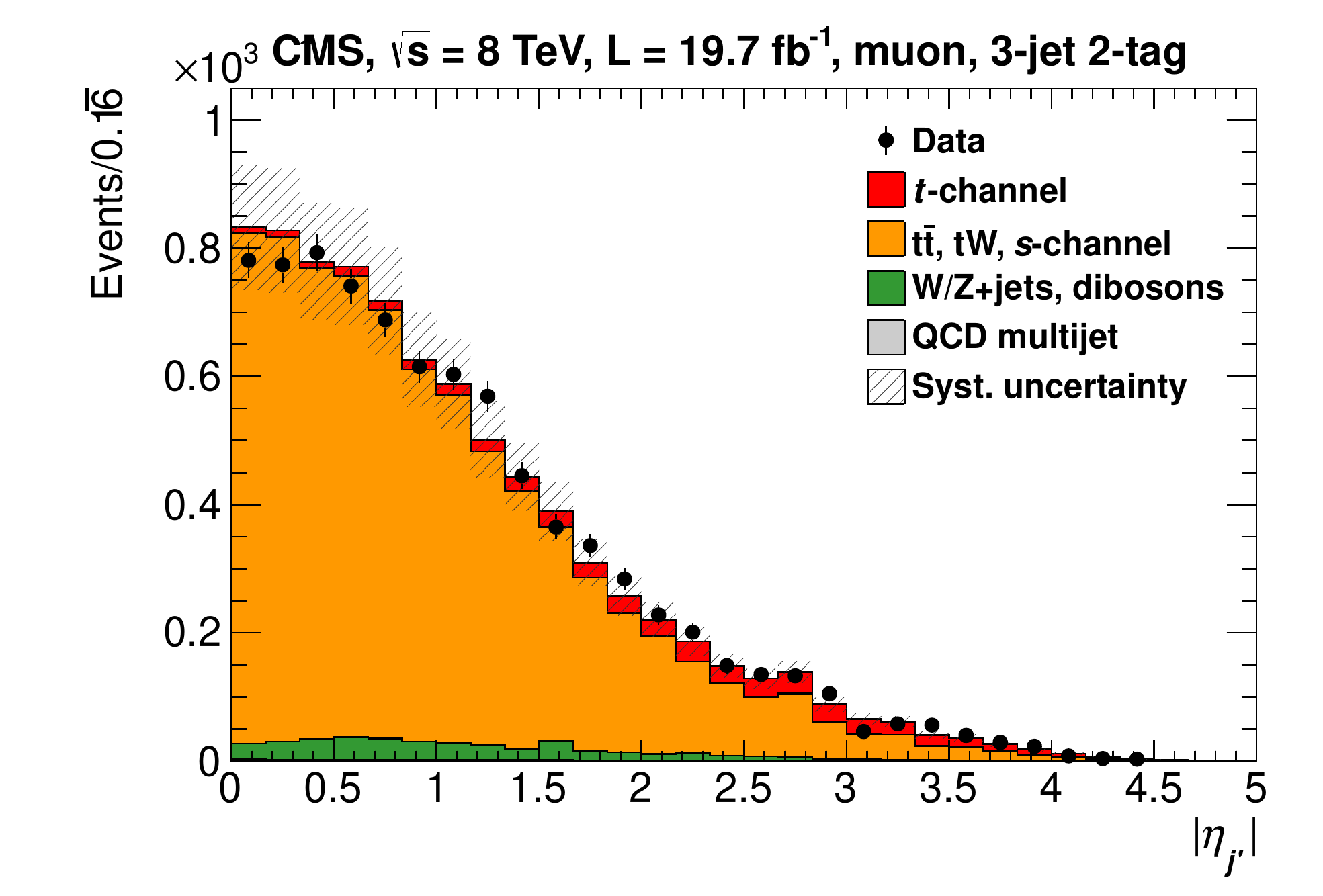}
\includegraphics[width=\cmsFigWidth]{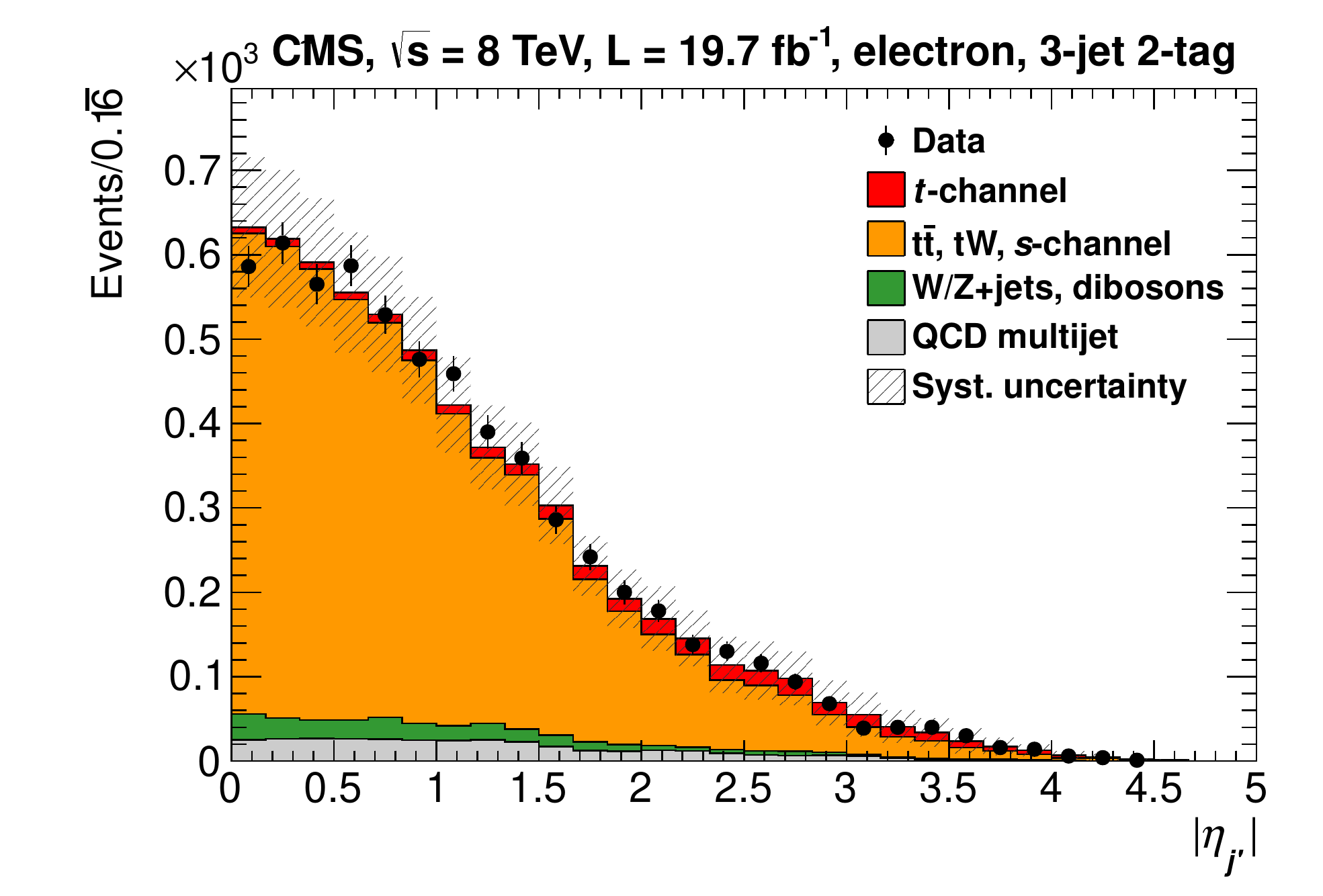}\\
\caption{\label{fig:ttbar_eta} Distribution of $\absetalj$ in the 3-jet 1-tag (upper left, upper right), and 3-jet 2-tag (lower left, lower right) samples for muon and electron decay channels.
The yield of the simulated processes is normalised to the results of the fit described in section~\ref{sec:etalj}. Systematic uncertainty
bands include all uncertainties.}
\end{figure}

\begin{figure}[!ht]
\centering
\includegraphics[width=\cmsFigWidth]{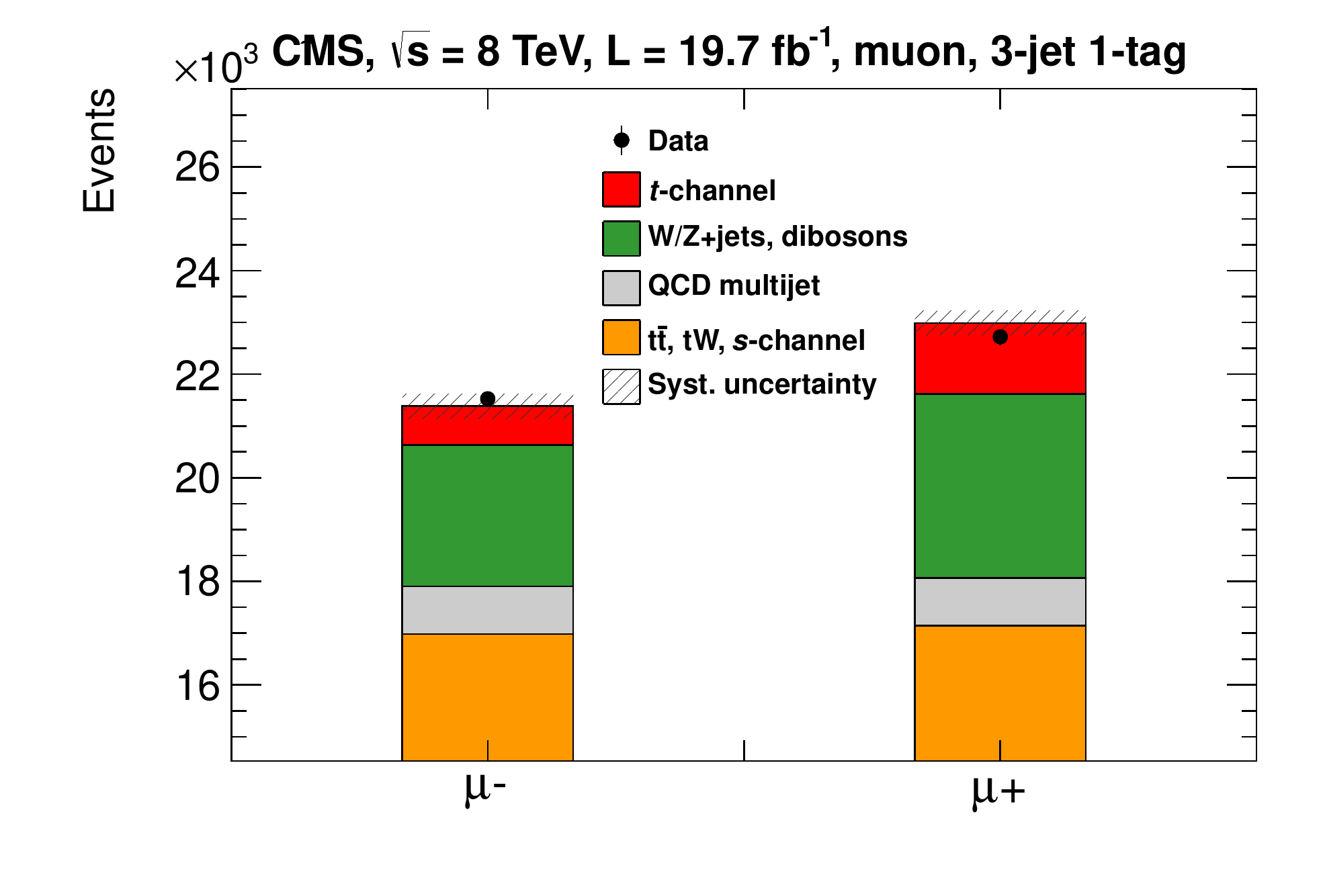}
\includegraphics[width=\cmsFigWidth]{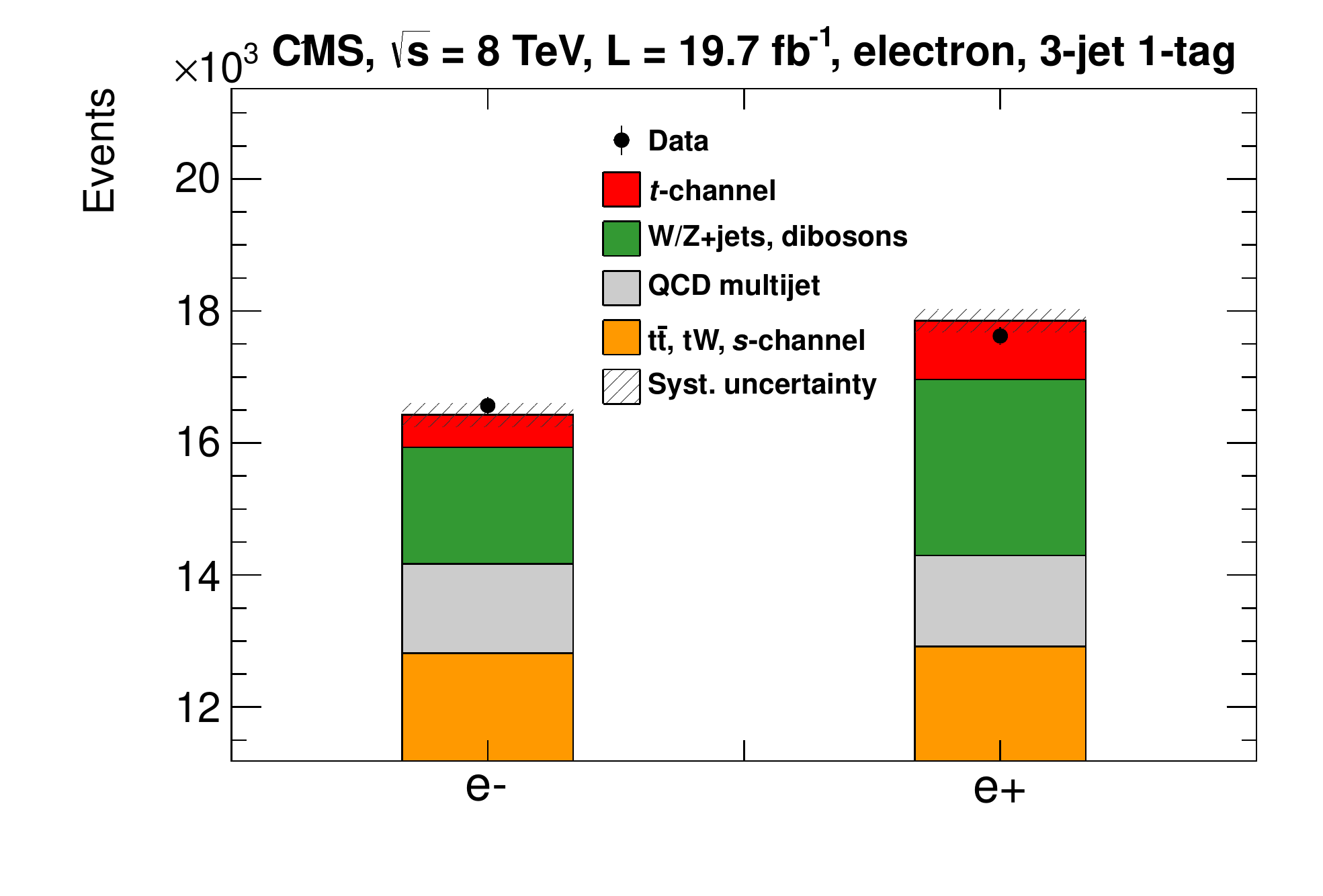}\\\
\includegraphics[width=\cmsFigWidth]{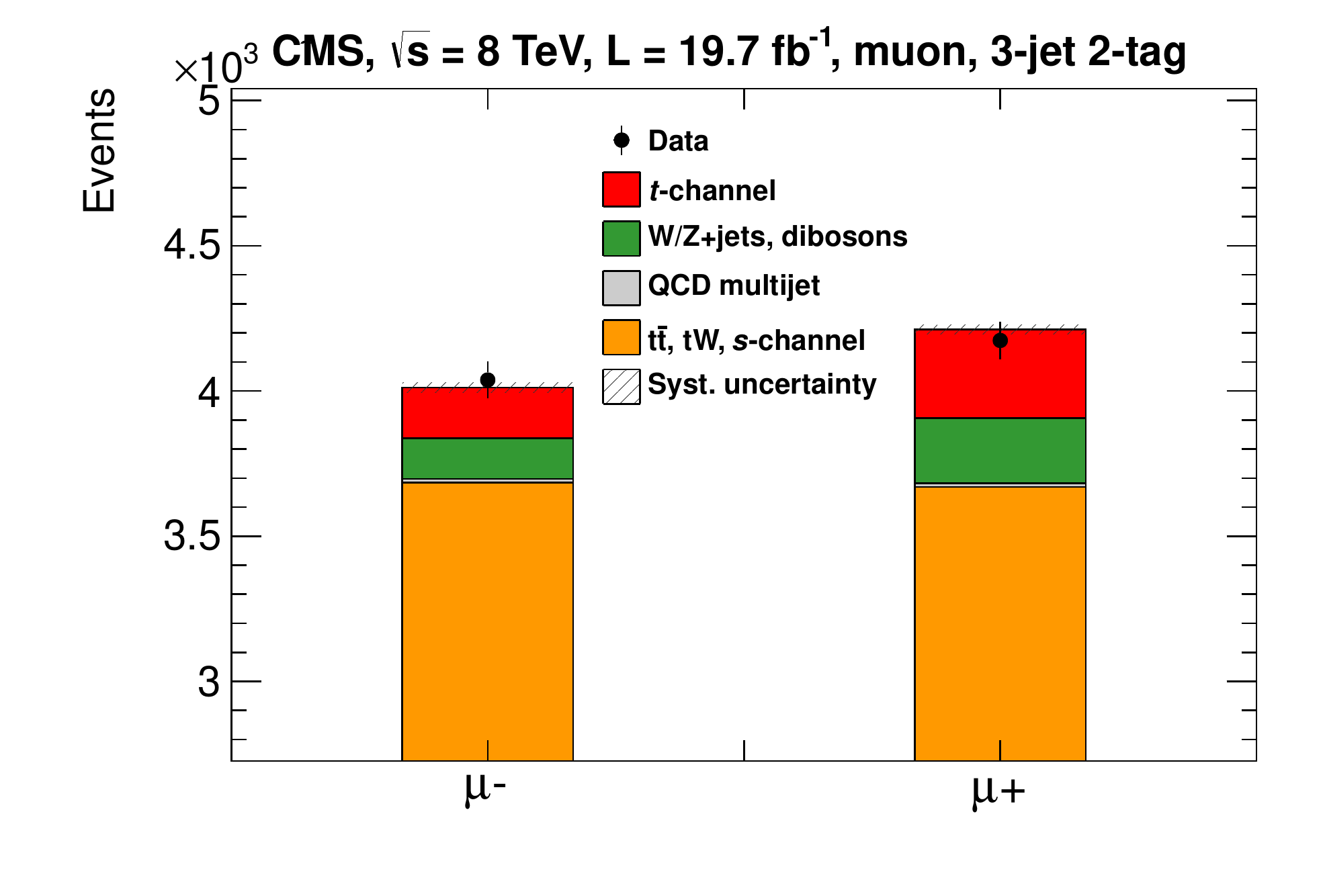}
\includegraphics[width=\cmsFigWidth]{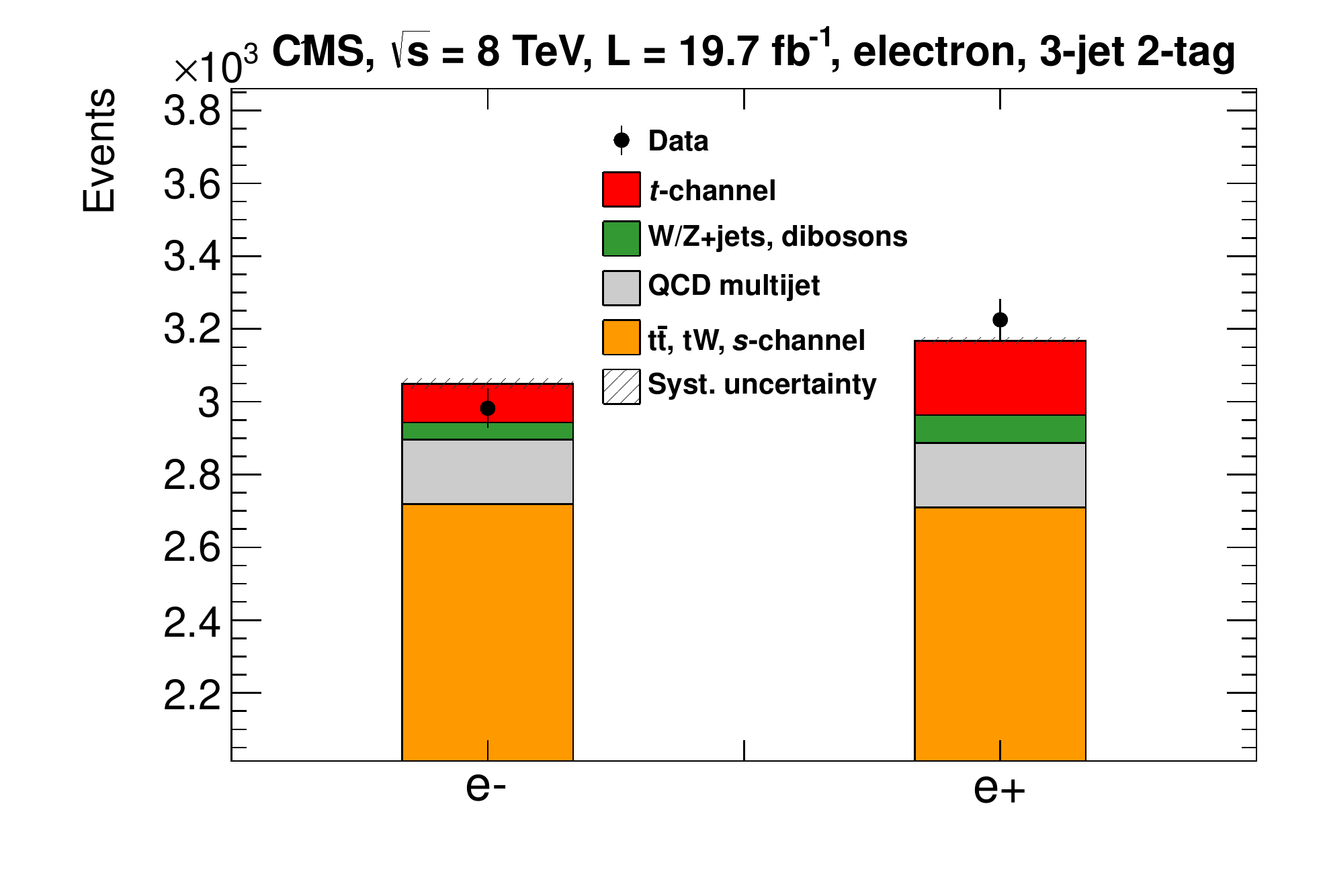}\\
\caption{\label{fig:ttbar_charge} Charge of the lepton in the 3-jet 1-tag (upper left, upper right), 3-jet 2-tag (lower left, lower right) samples for muon and electron decay channels. The sum of all predictions is normalised to the data yield. Systematic uncertainty bands include all uncertainties on the charge ratio.}
\end{figure}

\begin{figure}[!ht]
\centering
\includegraphics[width=\cmsFigWidth]{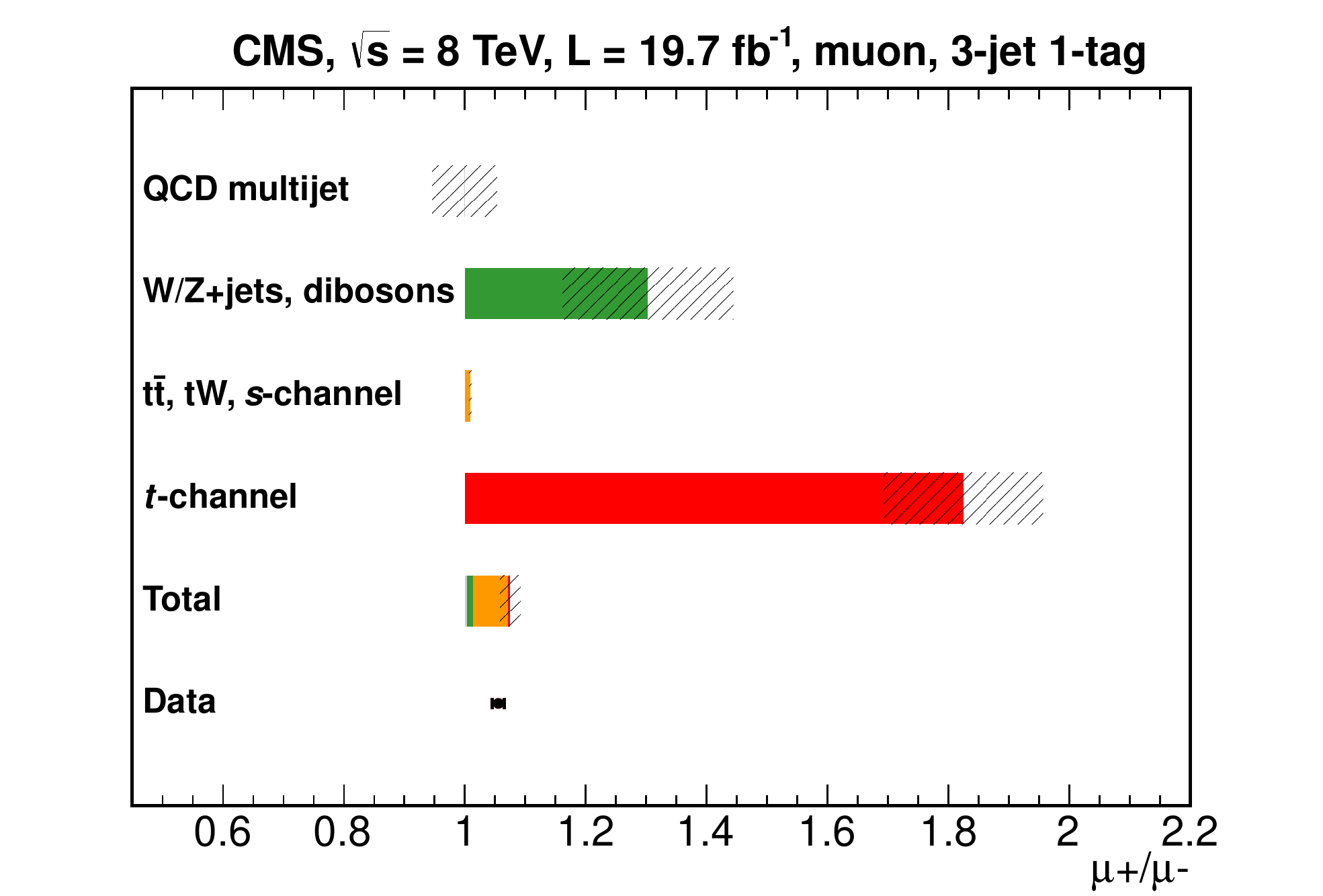}
\includegraphics[width=\cmsFigWidth]{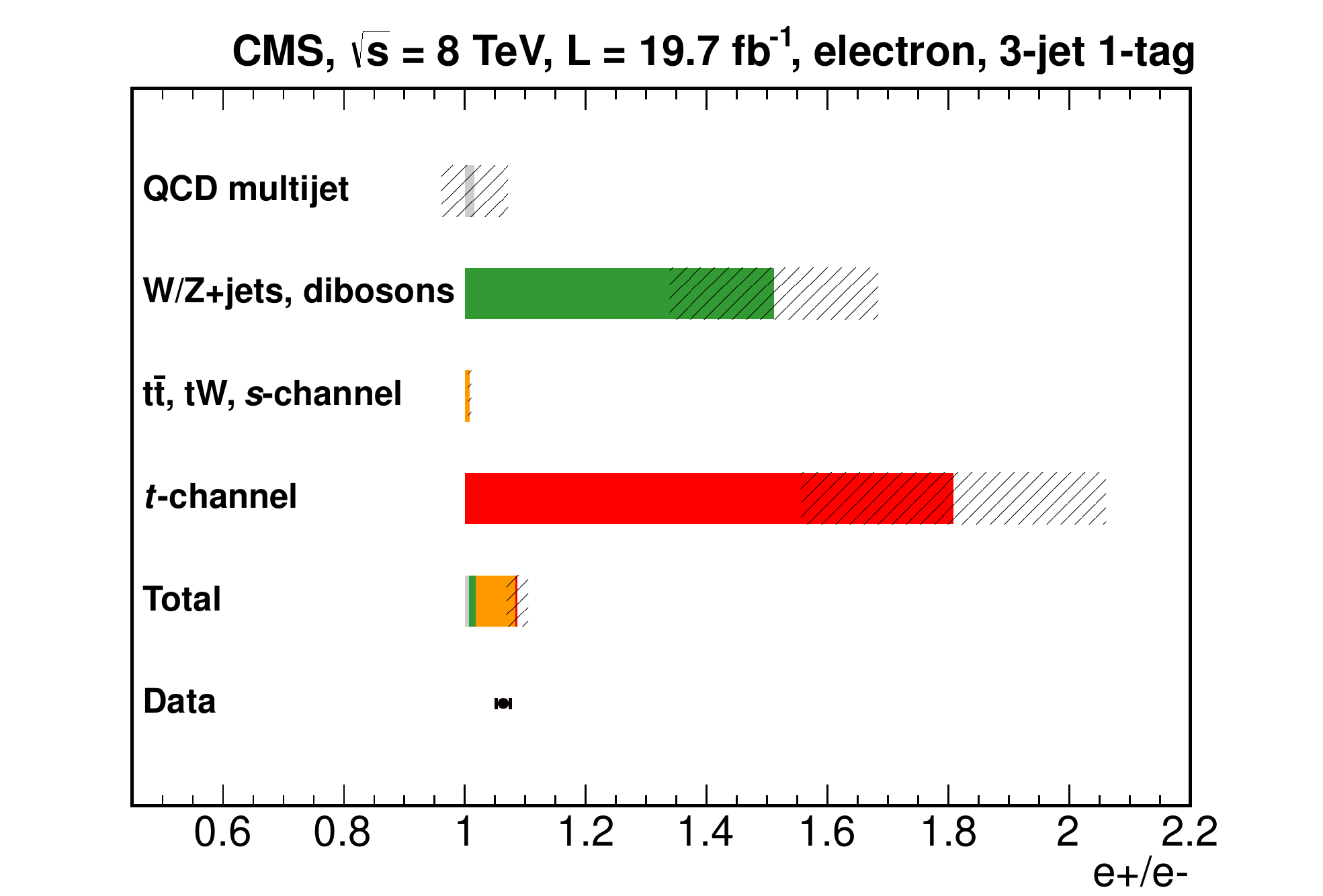}\\
\includegraphics[width=\cmsFigWidth]{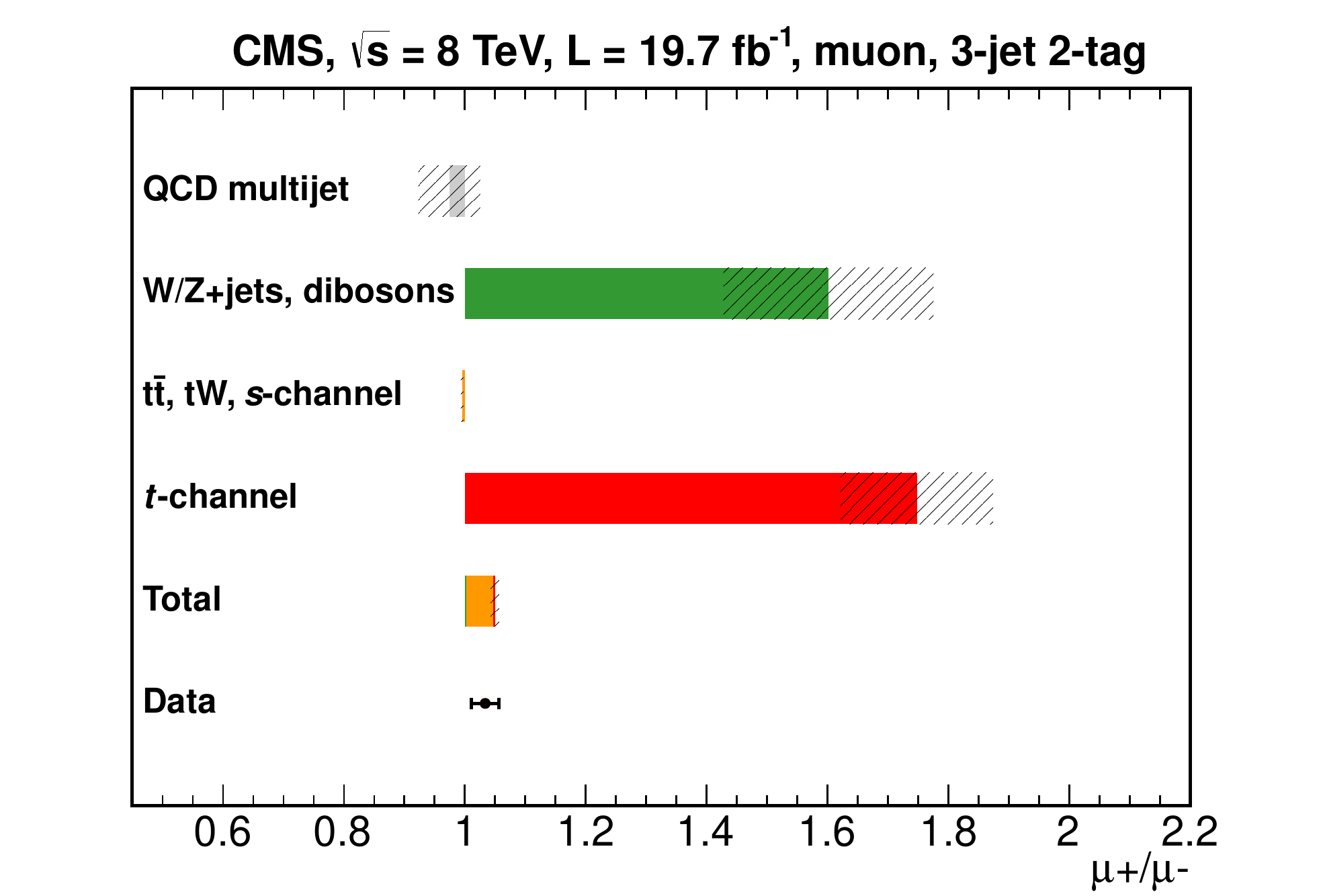}
\includegraphics[width=\cmsFigWidth]{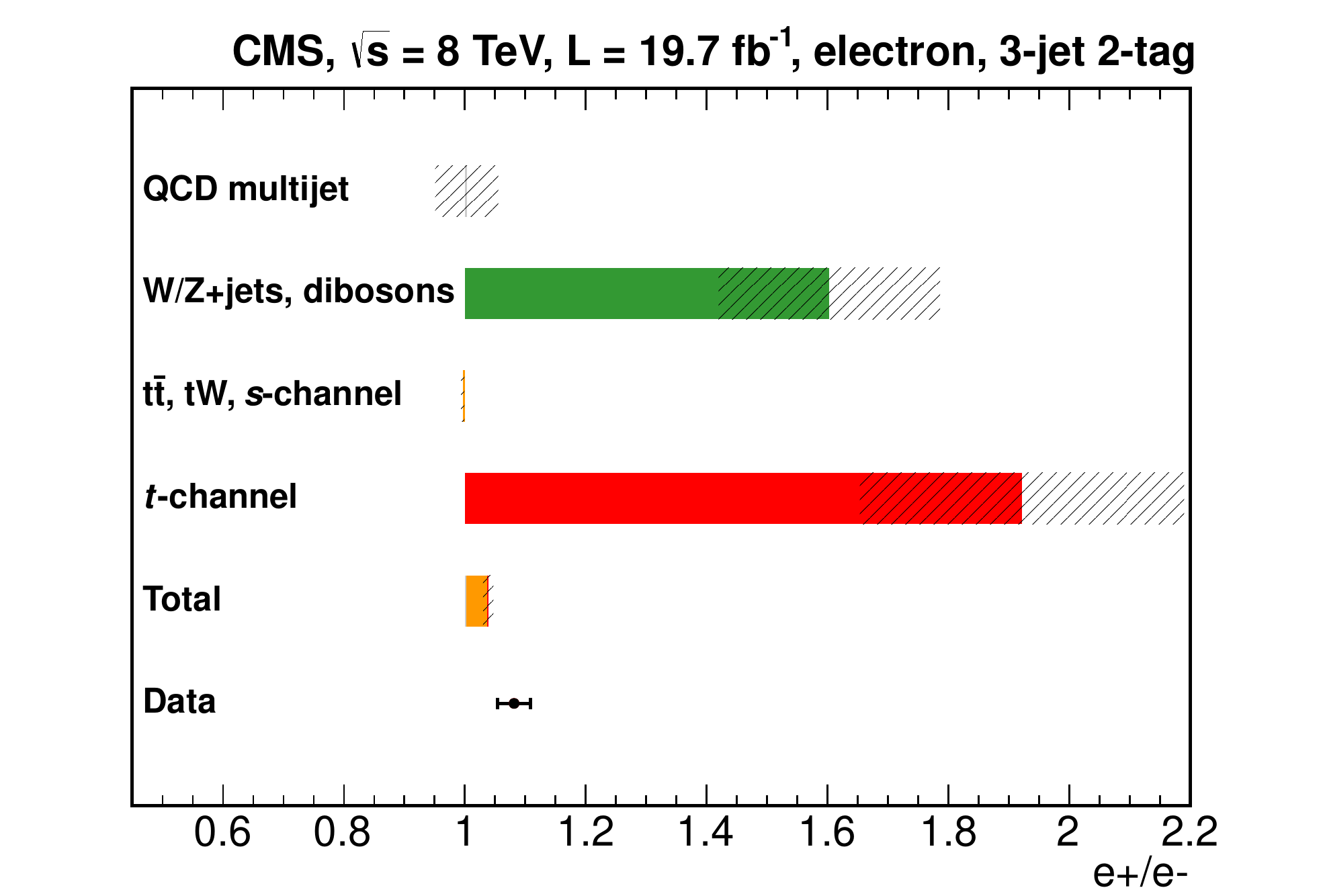}\\

\caption{\label{fig:ttbar_charge_ratio} Charge ratio between positively and negatively charged leptons in the 3-jet 1-tag (upper left, upper right), 3-jet 2-tag (lower left, lower right) samples for muon and electron decay channels. The charge ratio is shown separately for each process, as well as after normalising the sum of all predictions to the data yield. Systematic uncertainty bands include all uncertainties.}
\end{figure}

To reduce the dependence of the measurements on the modelling of \ttbar processes,
the $\absetalj$ distribution (template) used for signal extraction is modified
taking into account the $\absetalj$ distribution of the non-b-tagged jet in the 3-jet 2-tag sample
as follows.
The contribution of all SM processes except for \ttbar in the 3-jet 2-tag is subtracted from the template $\abs{\eta}$
distribution of the non-b-tagged jet taken from
data. Then the bin-by-bin ratio of the resulting template distribution and the
corresponding distribution from the \ttbar process is taken as the $\absetalj$-dependent correction factor for the \ttbar
in the 2-jet 1-tag sample. This ratio is then applied to the simulated distribution of $\absetalj$ in the
SR and SB.
\subsection{ The \texorpdfstring{\wzjets}{W/Z+jets} background}
\label{sec:whfextraction}
The 2-jet 0-tag sample is enriched with \wzjets background and it is used to test the agreement between simulation and data on the
distributions used for the signal extraction procedures.
The distribution of $\absetalj$ in the 2-jet 0-tag is shown in figure~\ref{fig:wjets_eta}, and good agreement between data and simulation is displayed.
The lepton charge in the 2-jet 0-tag sample is shown in figure~\ref{fig:wjets_charge}. The characteristic imbalance in the production
of positively and negatively charged leptons in $\wjets$ events can be seen clearly in the data, and the corresponding charge ratio
is shown in figure~\ref{fig:wjets_charge_ratio}.
The jets in this sample mostly originate from light quarks (u, d, s) or gluons, which tend to behave differently from
heavy-flavour jets (stemming from c and b quarks). For this reason, in the final fit procedure described later on
in section~\ref{sec:etalj} the $\wjets$ charge ratio is extracted from data as well.

\begin{figure}[!ht]
\centering
\includegraphics[width=\cmsFigWidth]{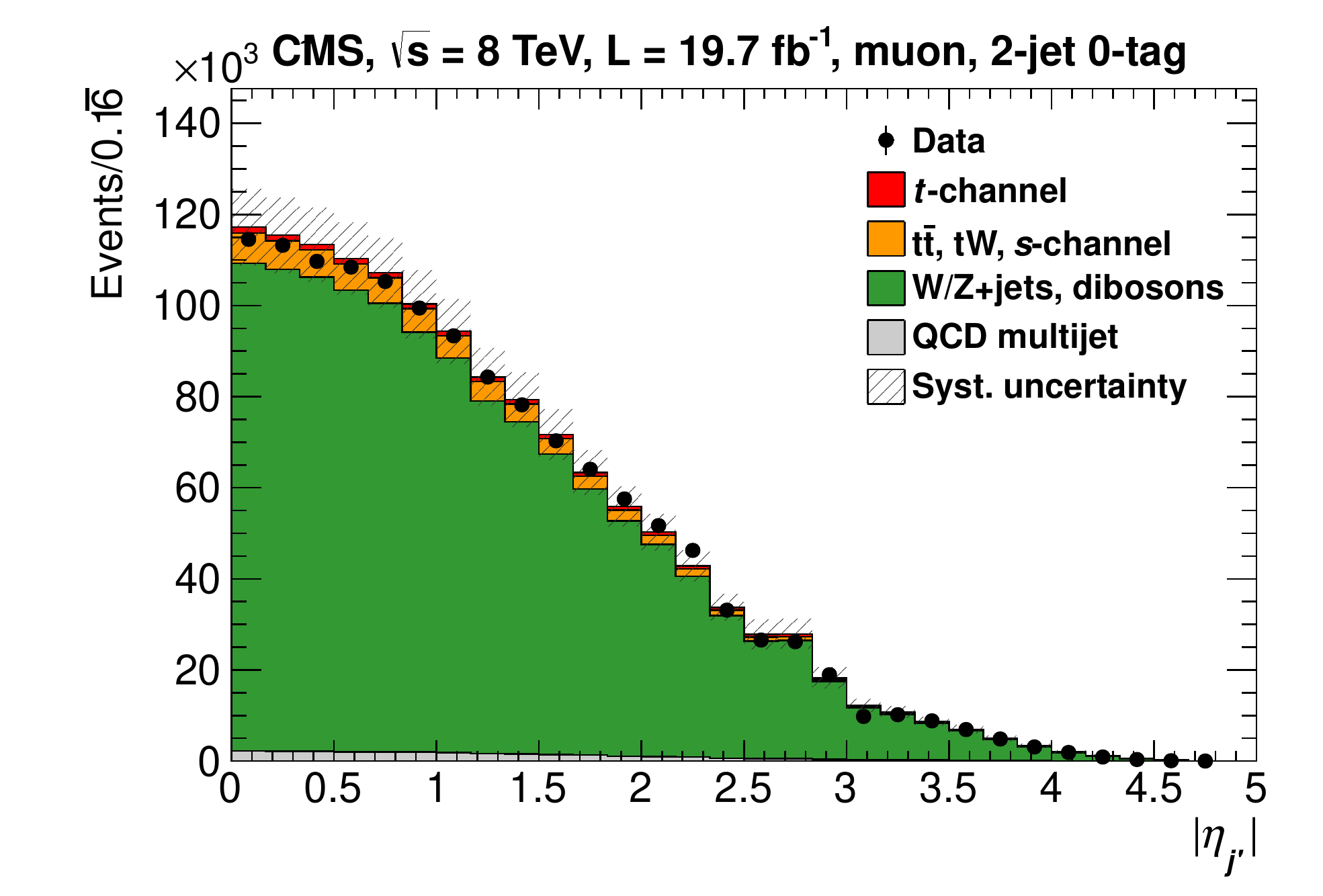}
\includegraphics[width=\cmsFigWidth]{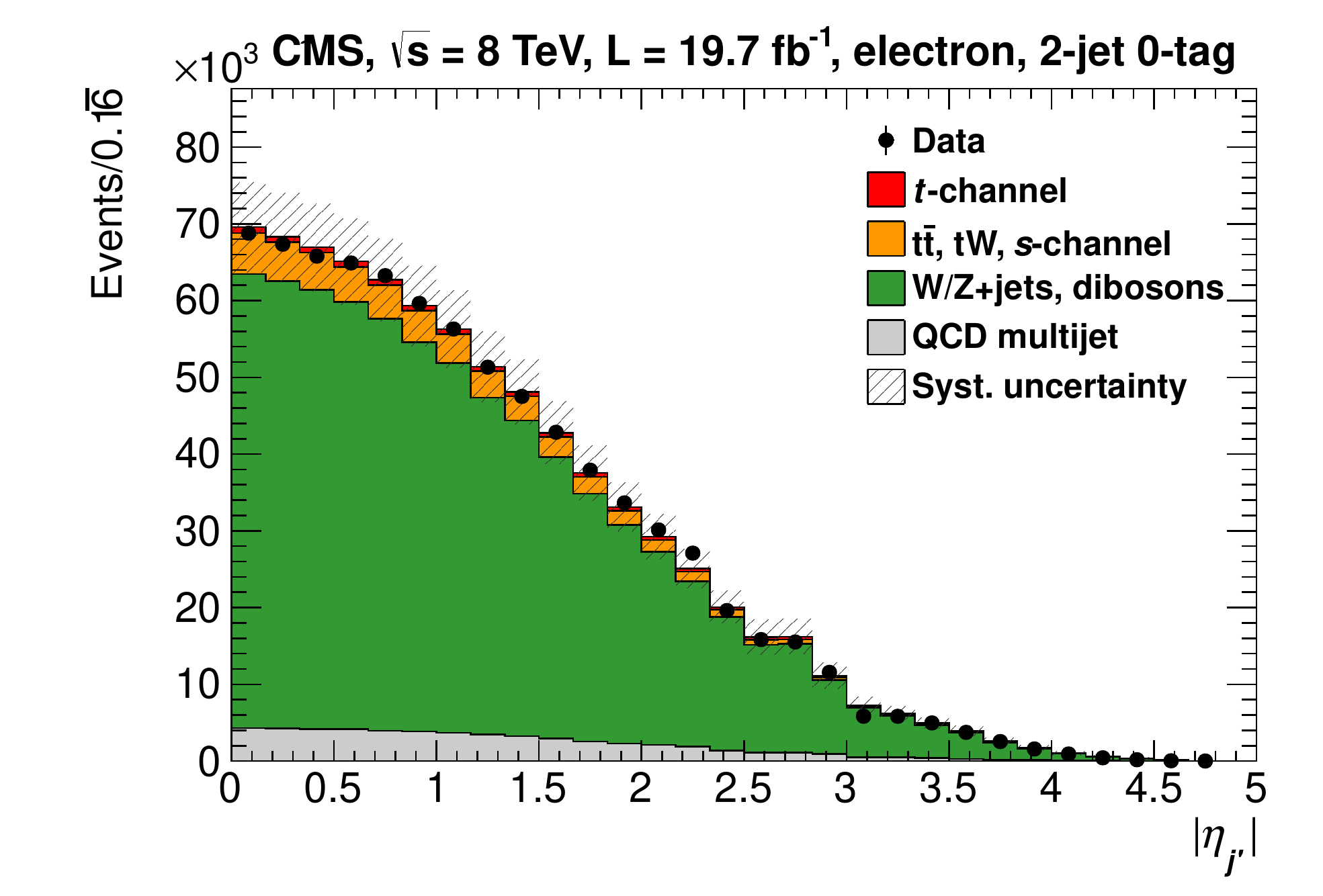}\\
\caption{\label{fig:wjets_eta} Distribution of $\absetalj$ in the 2-jet 0-tag sample for muon (left) and electron (right) decay channels. The \QCD contribution is derived from the fit to $\mTW$ and $\ETslash$. Systematic uncertainty bands include pre-fit uncertainties, both on the normalisation and on the shape of the distributions. }
\end{figure}

The SB region in the 2-jet 1-tag sample is used in order to estimate the \wzjets component in a region that is expected to have a similar composition in terms of \wzheavy flavours with respect to the sample that is used for the cross section extraction, i.e. the 2-jet 1-tag SR.
The $\absetalj$ distribution for \wzjets processes is taken from the sideband region by subtracting all other processes bin by bin. For this subtraction all samples except for  \ttbar and \QCD are derived from simulation. The latter two are estimated with the techniques described above. The scale factors between sideband region and signal region are derived from simulation. This procedure is performed for the inclusive distribution, as well as for positively and negatively charged leptons separately.
The bias due to the different kinematic properties of the two regions is estimated on simulations and removed, and the uncertainty on the composition in terms of W+c-jets and W+b-jets events is taken into account as described in section~\ref{sec:systs}.

\begin{figure}[!ht]
\centering
\includegraphics[width=\cmsFigWidth]{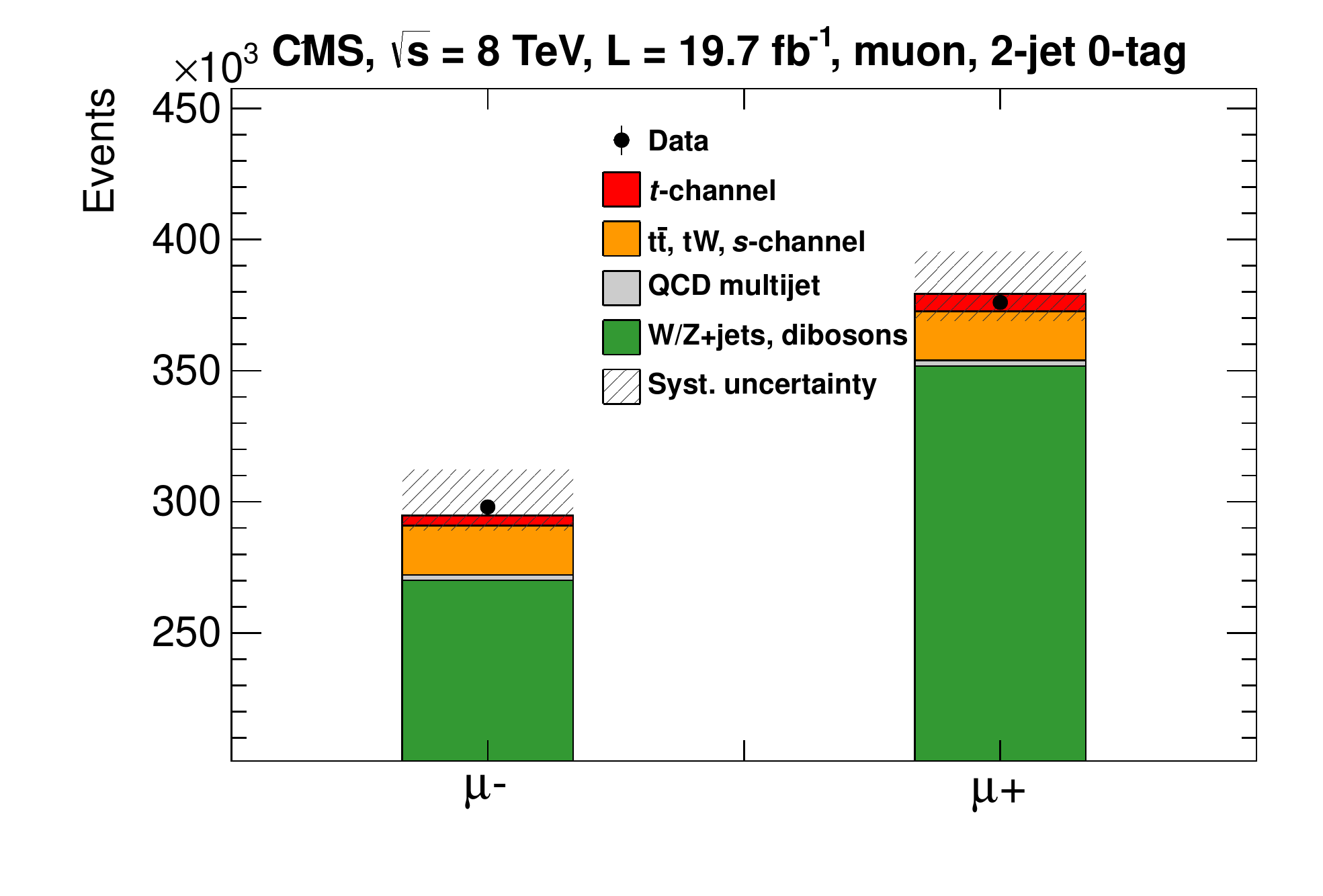}
\includegraphics[width=\cmsFigWidth]{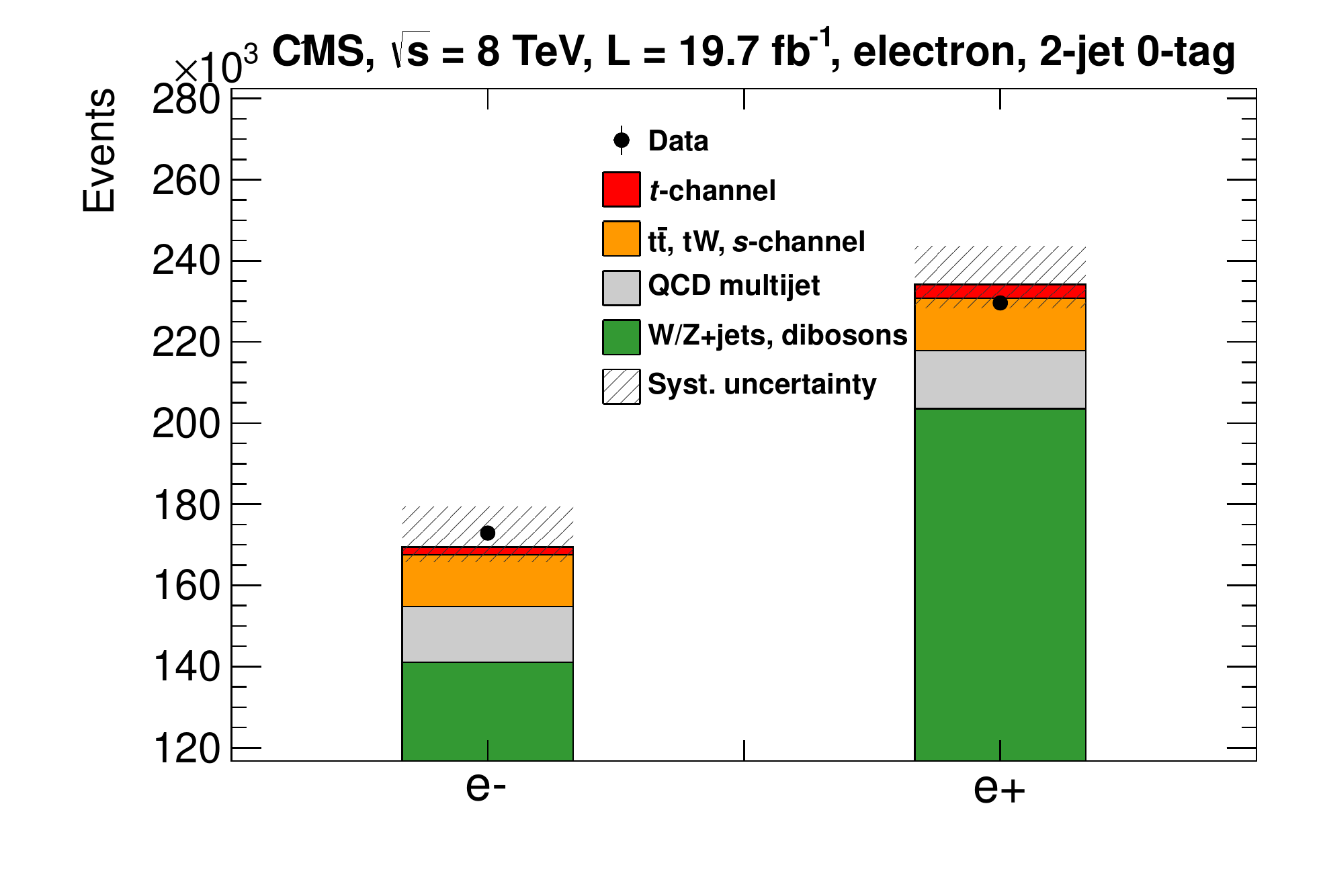}\\
\caption{ \label{fig:wjets_charge} Charge of the lepton in the 2-jet 0-tag sample for muon (left) and electron (right) decay channels. The sum of all predictions is normalised to the data yield. Systematic uncertainty bands include all uncertainties on the charge ratio.}
\end{figure}

\begin{figure}[!ht]
\centering
\includegraphics[width=\cmsFigWidth]{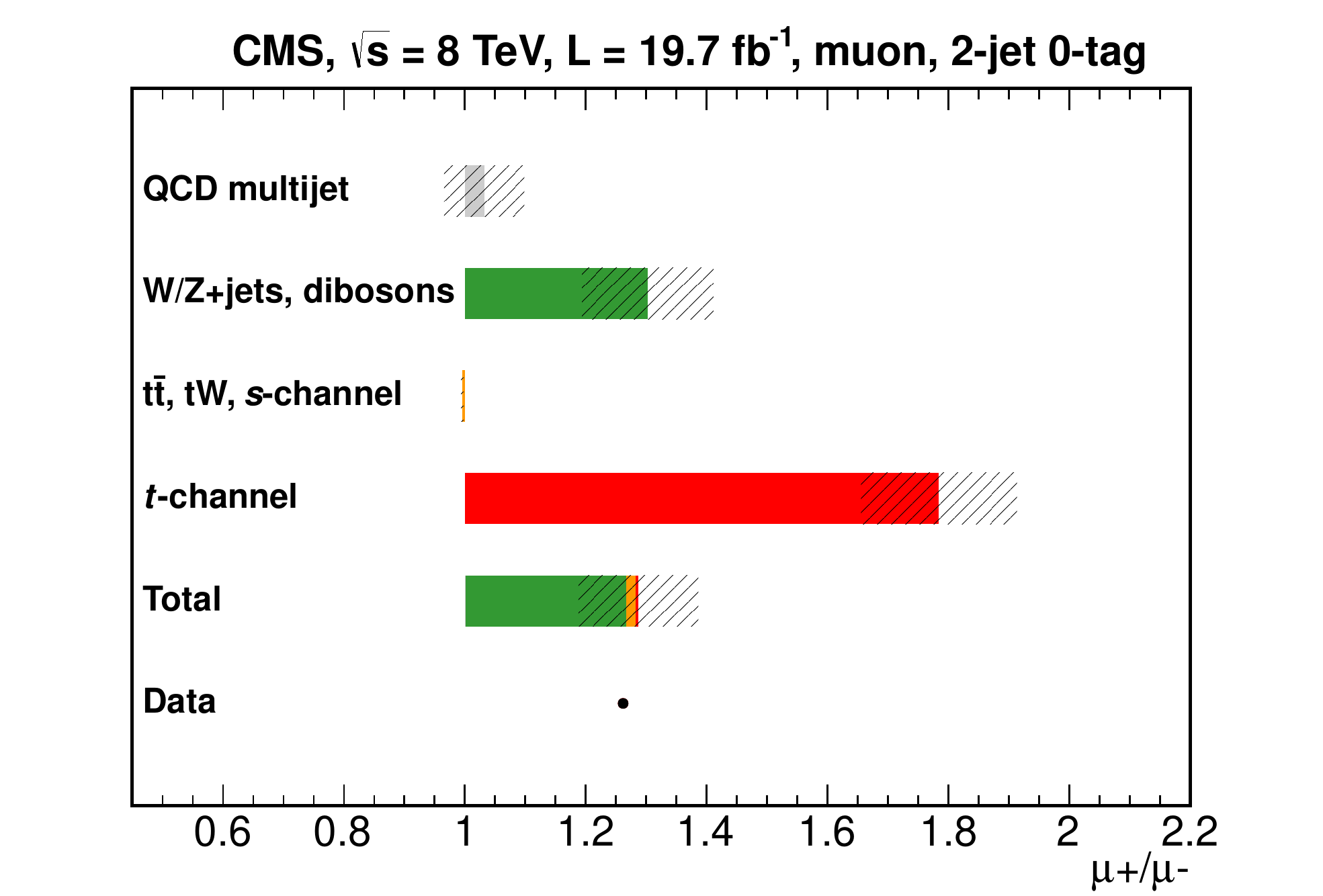}
\includegraphics[width=\cmsFigWidth]{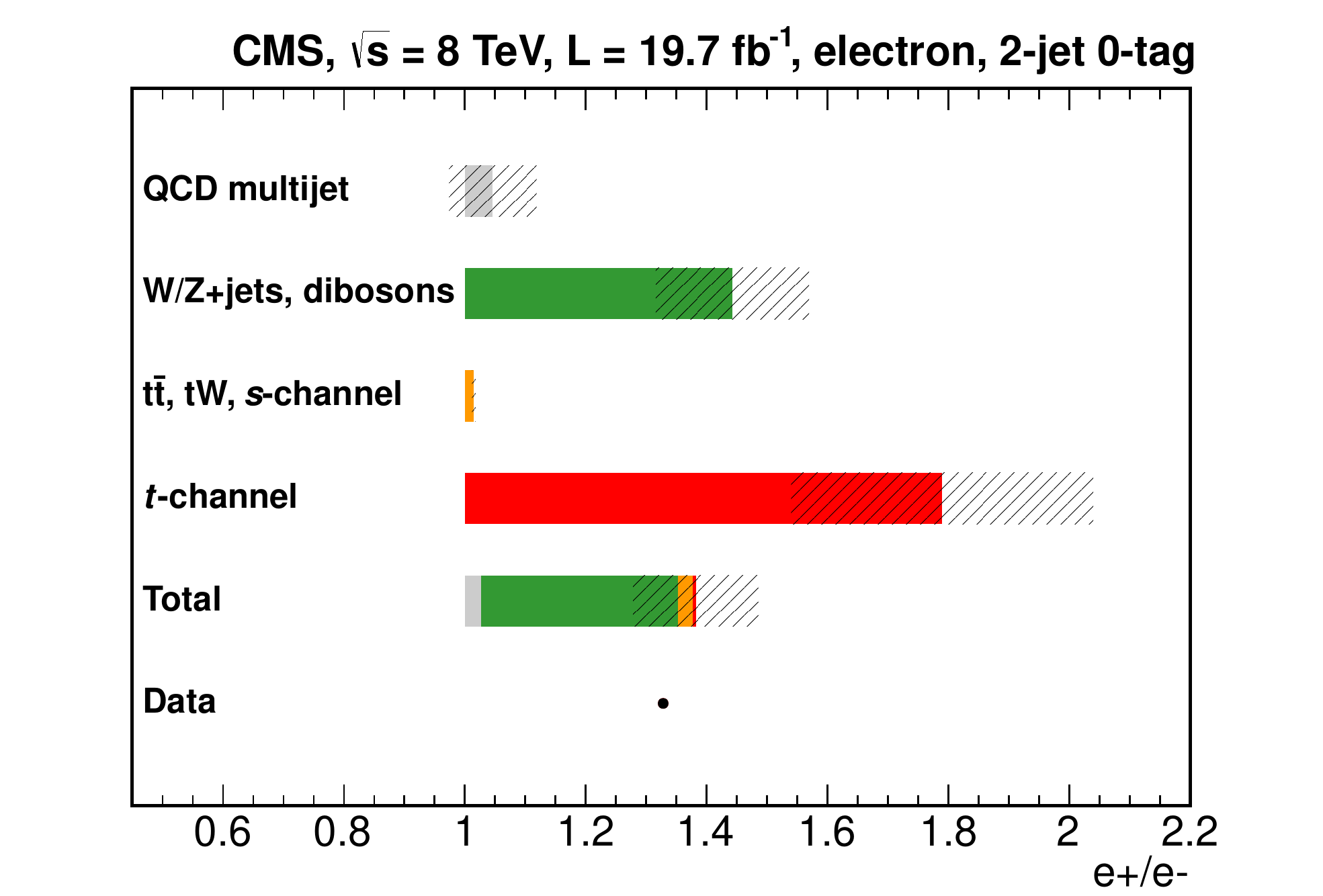}\\
\caption{\label{fig:wjets_charge_ratio} Charge ratio between positively and negatively charged leptons in the 2-jet 0-tag sample for muon (left) and electron (right) decay channels. The charge ratio is shown separately for each process, as well as after normalising the sum of all predictions to the data yield. Systematic uncertainty bands include all uncertainties.}
\end{figure}

\section{Signal extraction and cross section measurement}
\label{sec:etalj}
Two binned maximum-likelihood fits to the $\absetalj$ distributions of the events in the 2-jet 1-tag SR are performed.
The first fit extracts the inclusive single-top-quark cross section, the second extracts the separate single \tq and \tqbar cross sections.

The expected number of events in each $\absetalj$ bin is modelled with the following likelihood function:
\begin{equation}
\label{eq:likelihood}
n(\absetalj) =  N_s P_s(\absetalj) + N_{\cPqt} P_{\cPqt}(\absetalj) + N_\mathrm{EW} P_\mathrm{EW}(\absetalj) + N_\mathrm{MJ} P_\mathrm{MJ}(\absetalj).
\end{equation}

In addition to the signal (indicated with subscript $s$), three background components are considered:
the electroweak component (with subscript EW composed of W/Z+jets and dibosons), the top quark
component (with subscript t composed of \ttbar and single-top-quark tW and $s$-channel processes), and the QCD multijet component (with subscript MJ).
In equation~\ref{eq:likelihood},
$N_s$, $N_{\mathrm{EW}}$, $N_{\cPqt}$ and $N_{\mathrm{MJ}}$ are the yields of the signal and of the three background components;
$P_{\mathrm{s}}$, $P_{\mathrm{b}}$ (b=EW, t, MJ) are the binned probability distribution functions for the signal and for the
different background components.

The inclusive cross section is extracted from events with positively or negatively charged leptons, defining one likelihood function per lepton flavour, as in equation~\ref{eq:likelihood}, then fitting simultaneously the two distributions for muons and electrons. The single \tq and \tqbar cross sections are extracted by further dividing the events by lepton charge, defining one likelihood function per lepton flavour and per charge, as in equation~\ref{eq:likelihood}, then fitting simultaneously the four distributions.

The definition of the probability distribution functions and of the parameters included in the fit are described in the following:
\begin{itemize}
\item \textbf{Signal:} $P_{\mathrm{s}}$ for both fits is taken from simulation (see also section~\ref{sec:datasets}) as the predicted $\absetalj$ distribution.
The total yield $N_s$ is fitted unconstrained in the inclusive single-top-quark cross section fit. Two parameters are introduced
in the single \tq and \tqbar cross section fit for the positively and negatively charged lepton signal yield and fitted unconstrained.
\item \textbf{EW component: W/Z + jets, diboson:} The $P_{\mathrm{EW}}$ distribution is taken as the sum of the contribution of $\Vjets$ and
diboson processes. The $\vjets$ normalisation and distribution are estimated from the $\mt$ sideband with the method described in section~\ref{sec:backgrounds}. This sideband method is applied to both muons and electrons, inclusively with respect to the
lepton charge in the case of the inclusive top-quark cross section fit, and separately for positively and negatively charged leptons in the case of the
single \tq and \tqbar cross section fit. The diboson contribution is then taken from simulation.
The two contributions are summed together and the total yield $N_{\mathrm{EW}}$ is derived by the fit.
To take into account the prior knowledge of the normalisation obtained from the sideband a Gaussian constraint is applied to $N_{\mathrm{EW}}$ in the fit, i.e. the likelihood function is further multiplied by a Gaussian function of $N_{\mathrm{EW}}$. The mean value of
this function is taken from the procedure previously described in this paragraph, while the standard deviation is taken equal to the difference between the data-based yield of $\VJets$ and the expectation from simulation in the sideband region.
For the single \tq and \tqbar cross section ratio fit, the $N_{\mathrm{EW}}$ are fitted separately for positively
and negatively charged leptons.
\item \textbf{Top quark component: \tt, tW and $s$-channel:} $P_{\cPqt}$ is taken from the data-based procedure described in section~\ref{sec:backgrounds}, to which the single-top-quark tW and $s$-channel processes are added with a normalisation factor taken from simulation. This contribution is separated by lepton flavour and charge assuming charge symmetry of $\tt$ and tW events. The $s$-channel charge ratio is fixed to the SM prediction.
The yield  $N_{\cPqt}$ is then fitted with a Gaussian constraint, centred on the value obtained from simulation and with a variation of $\pm$10\%, which is chosen to cover both experimental and theoretical uncertainties on the $\tt$ cross section.
\item \textbf{QCD multijet:} $P_{\mathrm{MJ}}$ is taken from the QCD multijet enriched sample defined in section~\ref{sec:backgrounds},
adding an extra requirement on the angular distance of the lepton and the jets, $\Delta R(\ell, \mathrm{j})> 0.3$.
The yield is fixed to the results of the $\mTW$ and $\ETslash$ fit.
\end{itemize}
The fit strategy driving this parametrisation is focused on constraining from data the $\wzjets$ and $\ttbar$
backgrounds. In the particular case of the single \tq and \tqbar cross section fit, the event ratio of positively
and negatively charged W bosons is constrained as well. The cross sections are extracted using the detector acceptance derived from the simulated signal sample.
The total cross section measurement from the inclusive analysis is more precise than the one inferred from the separate-by-charge fit, due to the additional uncertainty from the W charged ratio, which is extracted from data.
The $\absetalj$ distributions for the muon and electron decay channels obtained by normalising the contribution of each process to the value of the inclusive cross section and \tq and \tqbar cross section ratio fits are shown in figures~\ref{fig:fits_tot} and ~\ref{fig:fits_ch}, respectively.
\begin{figure}[!ht]
\centering
\includegraphics[width=\cmsFigWidth]{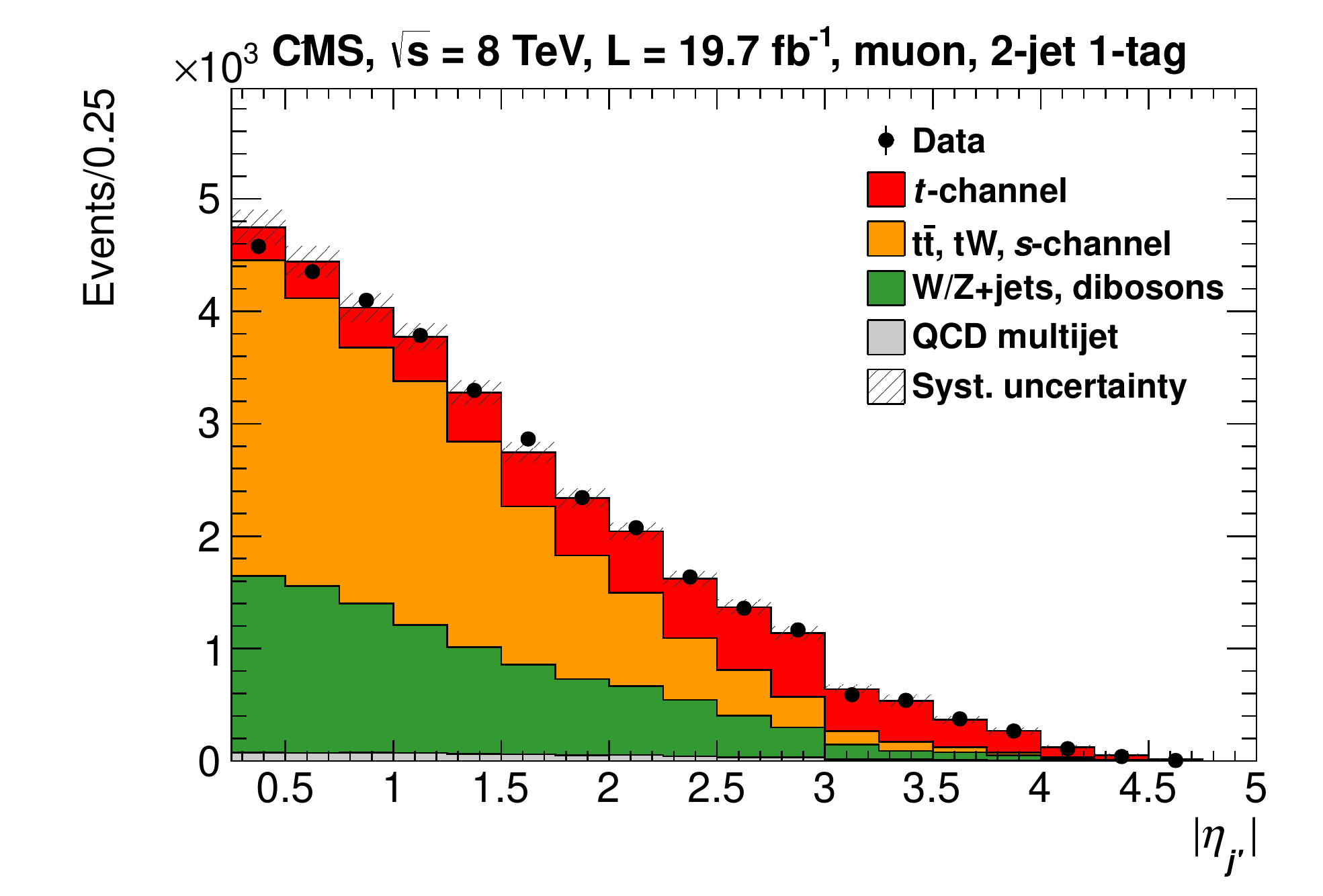}
\includegraphics[width=\cmsFigWidth]{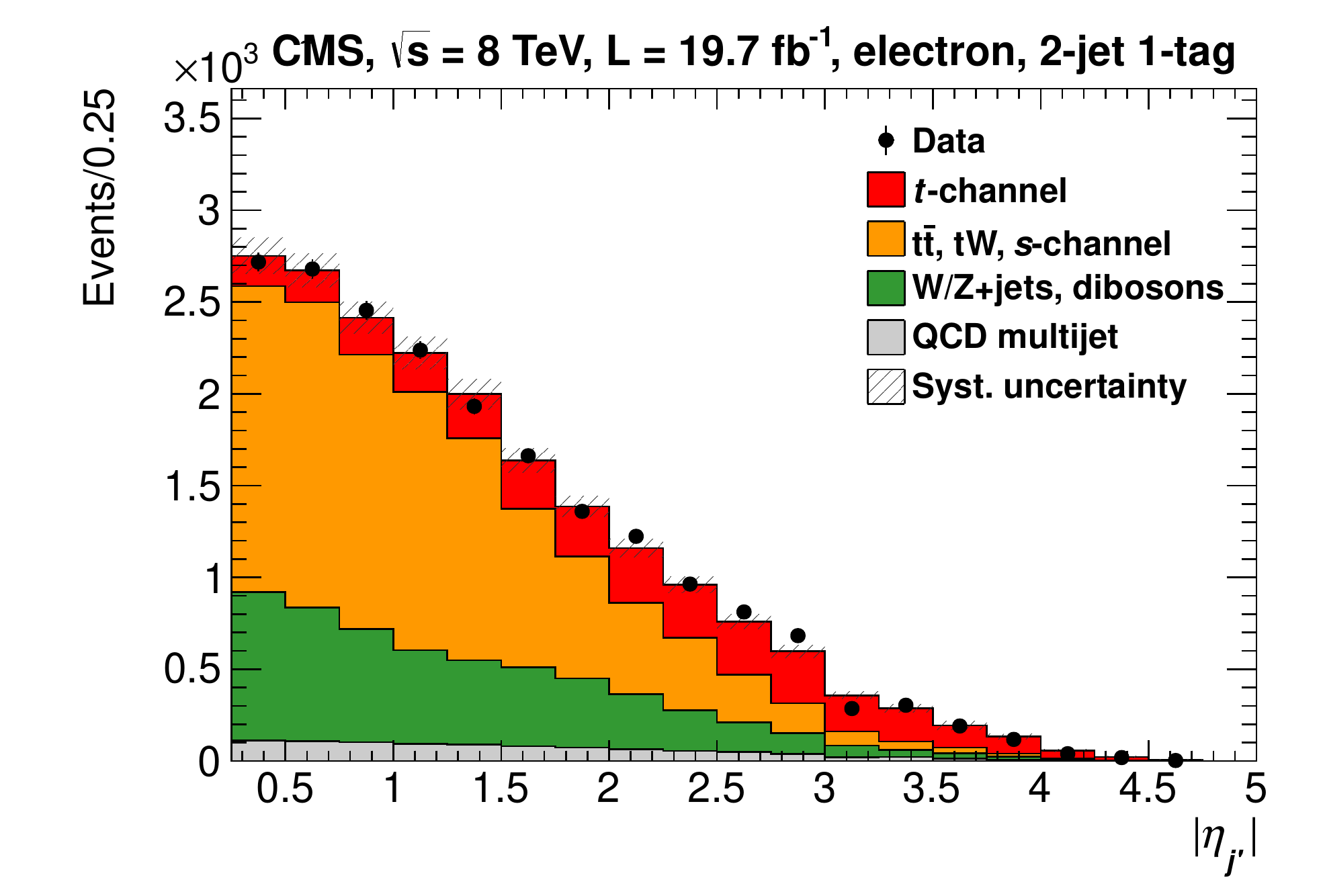}\\
\caption{\label{fig:fits_tot} Fitted $\absetalj$ distributions for muon (left) and electron (right) decay channels, normalised to the yields obtained from the combined total cross section fit. Systematic uncertainty bands include the shape uncertainties on the distributions. }
\end{figure}
\begin{figure}[!ht]
\centering
\includegraphics[width=\cmsFigWidth]{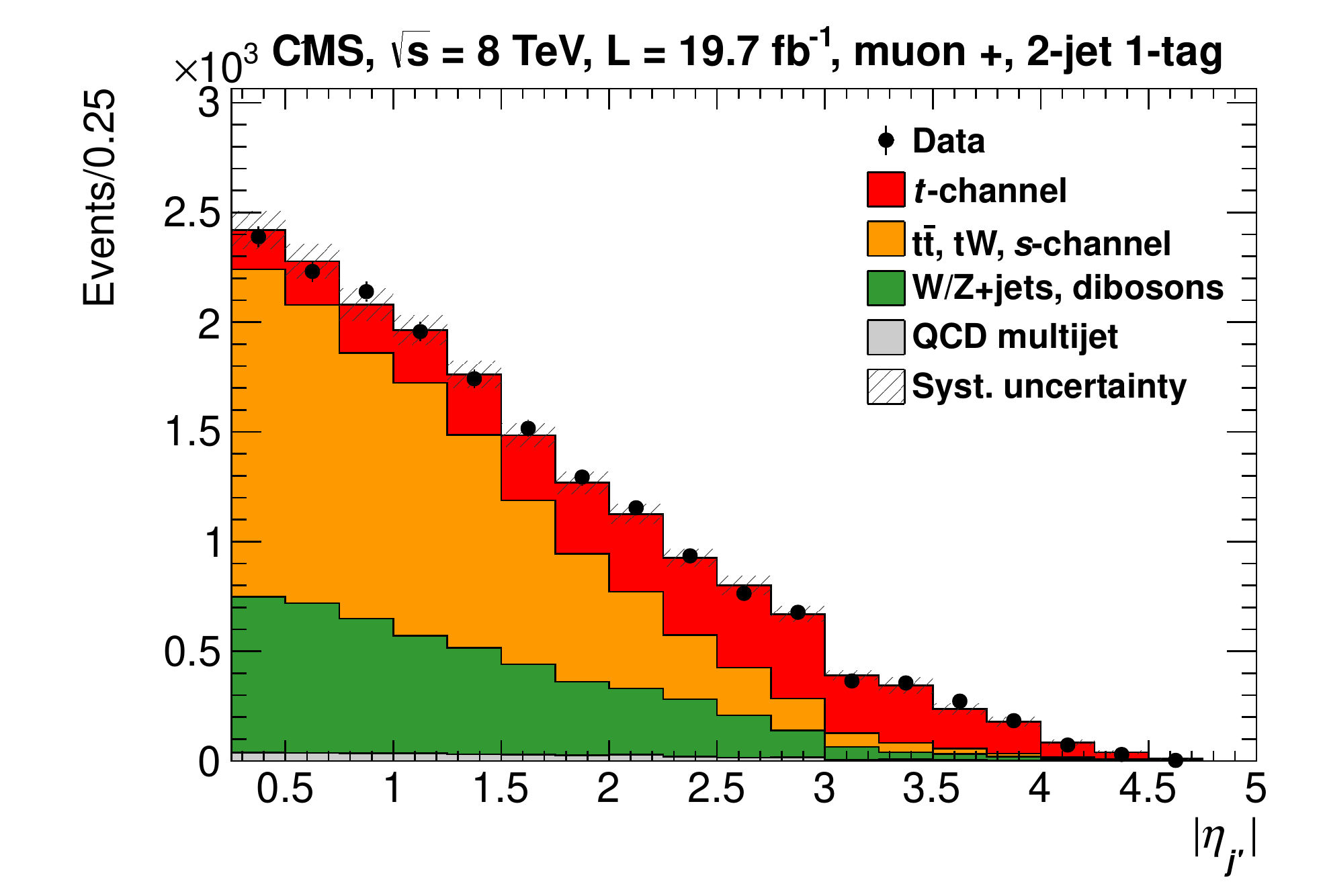}
\includegraphics[width=\cmsFigWidth]{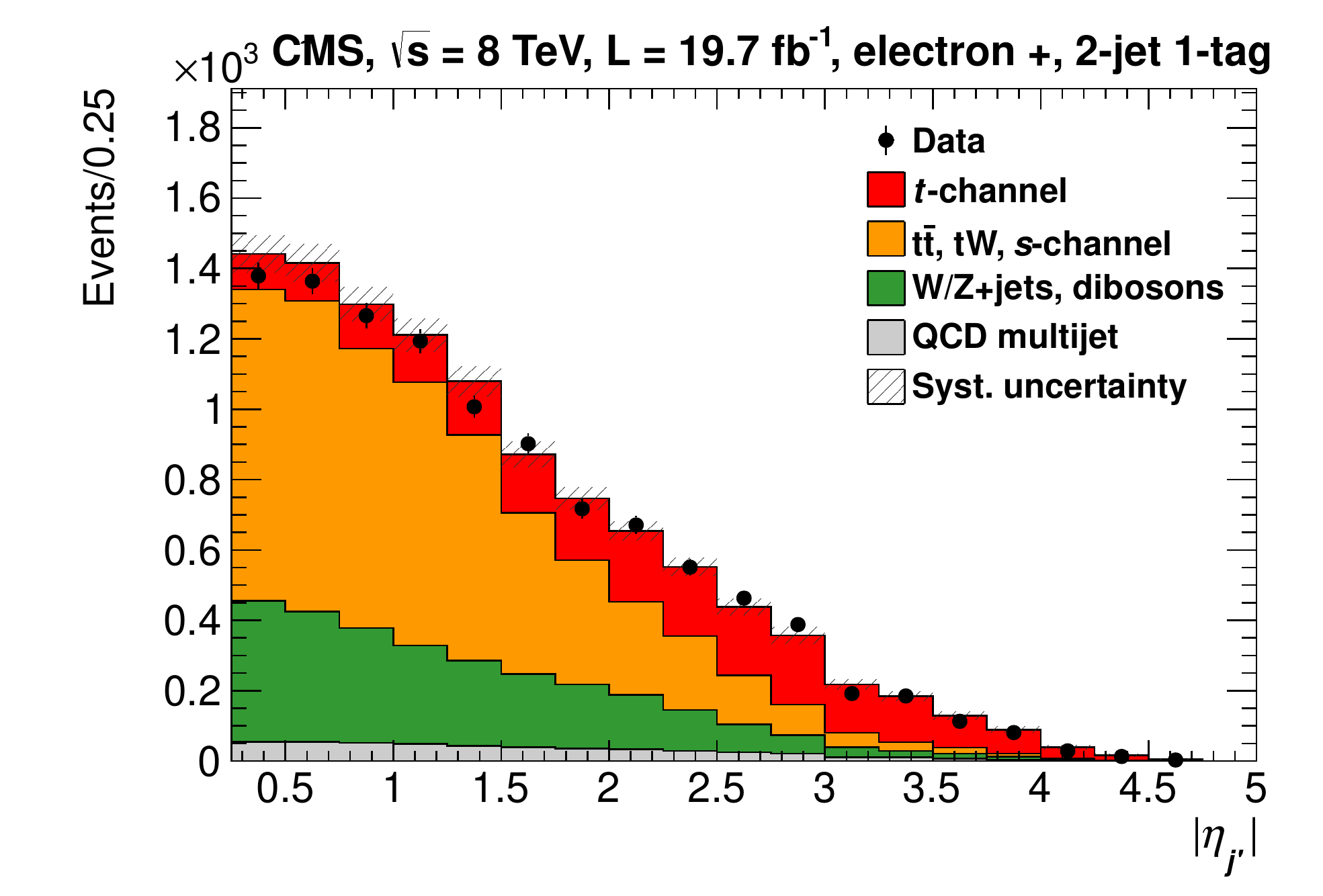}\\
\includegraphics[width=\cmsFigWidth]{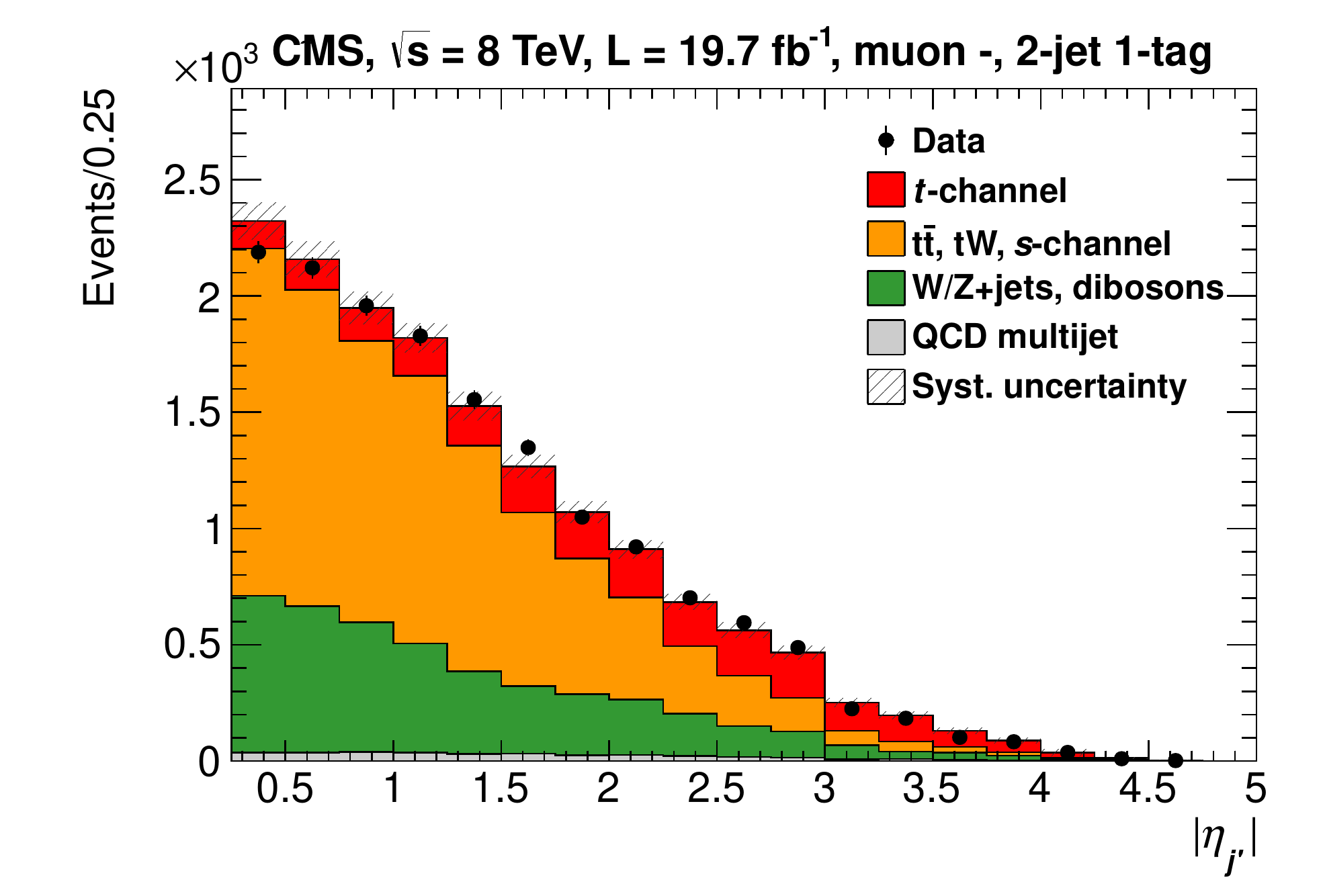}
\includegraphics[width=\cmsFigWidth]{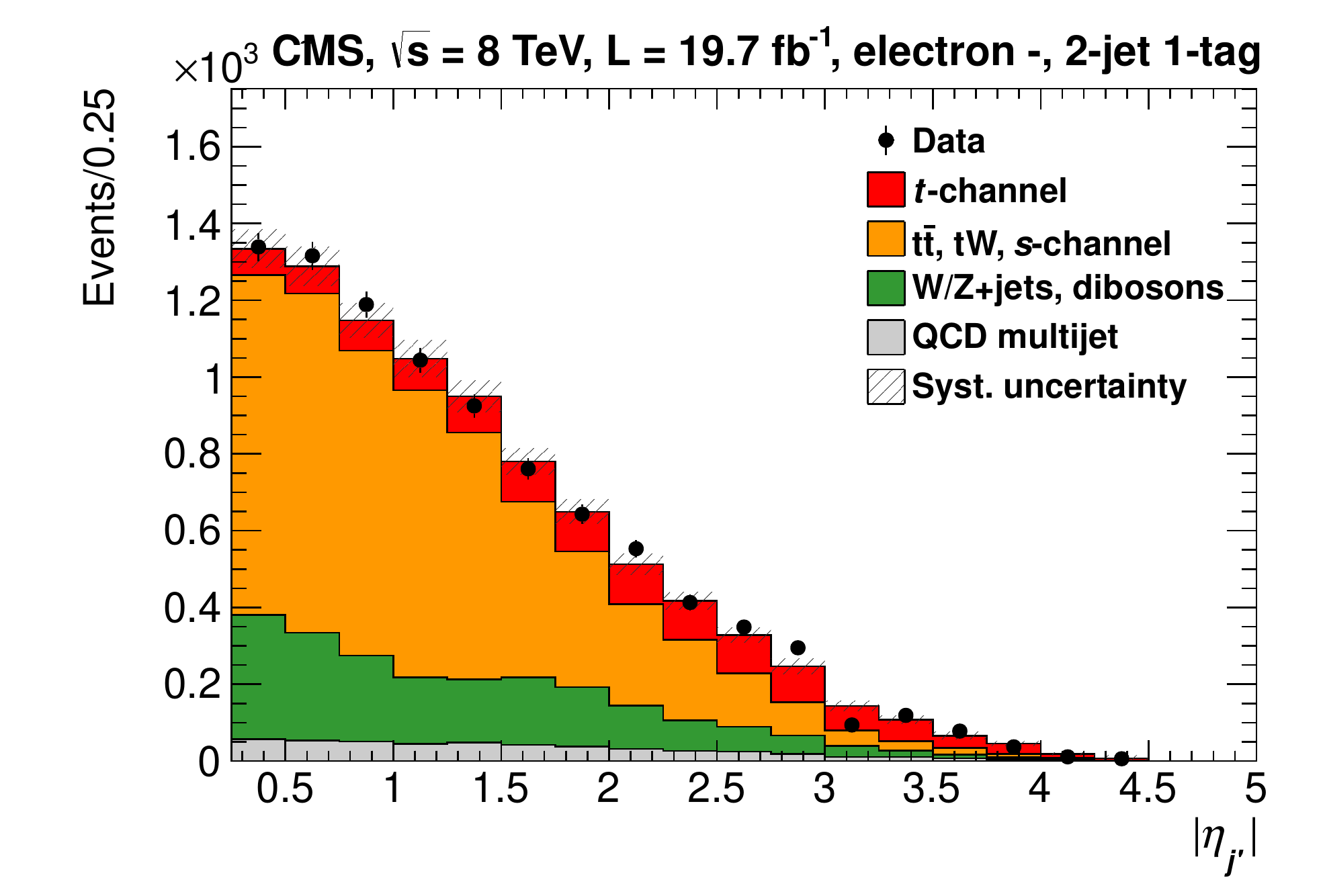}
\caption{\label{fig:fits_ch} Fitted $\absetalj$ distributions for muon (upper left, lower left) and electron (upper right, lower right) decay channels, normalised to the yields obtained from the combined single \tq and \tqbar cross section ratio fit. Systematic uncertainty bands include the shape uncertainties on the distributions.}
\end{figure}
An indication of the validity of the fit extraction procedure comes from the study of characteristic $t$-channel properties
in the signal sample after normalising each process to the fit results.
The reconstructed top-quark mass \mt in the region with $\absetalj>2.5$, after scaling each
process contribution to the normalisation obtained from the fit, is shown in figure~\ref{fig:fitsTopMass}.
This region is expected to be depleted of background events and enriched in $t$-channel signal events, hence displaying a characteristic peak around the top-quark mass value, which appears clearly in data for
both the muon and the electron channels.
\begin{figure}[!ht]
\centering
\includegraphics[width=\cmsFigWidth]{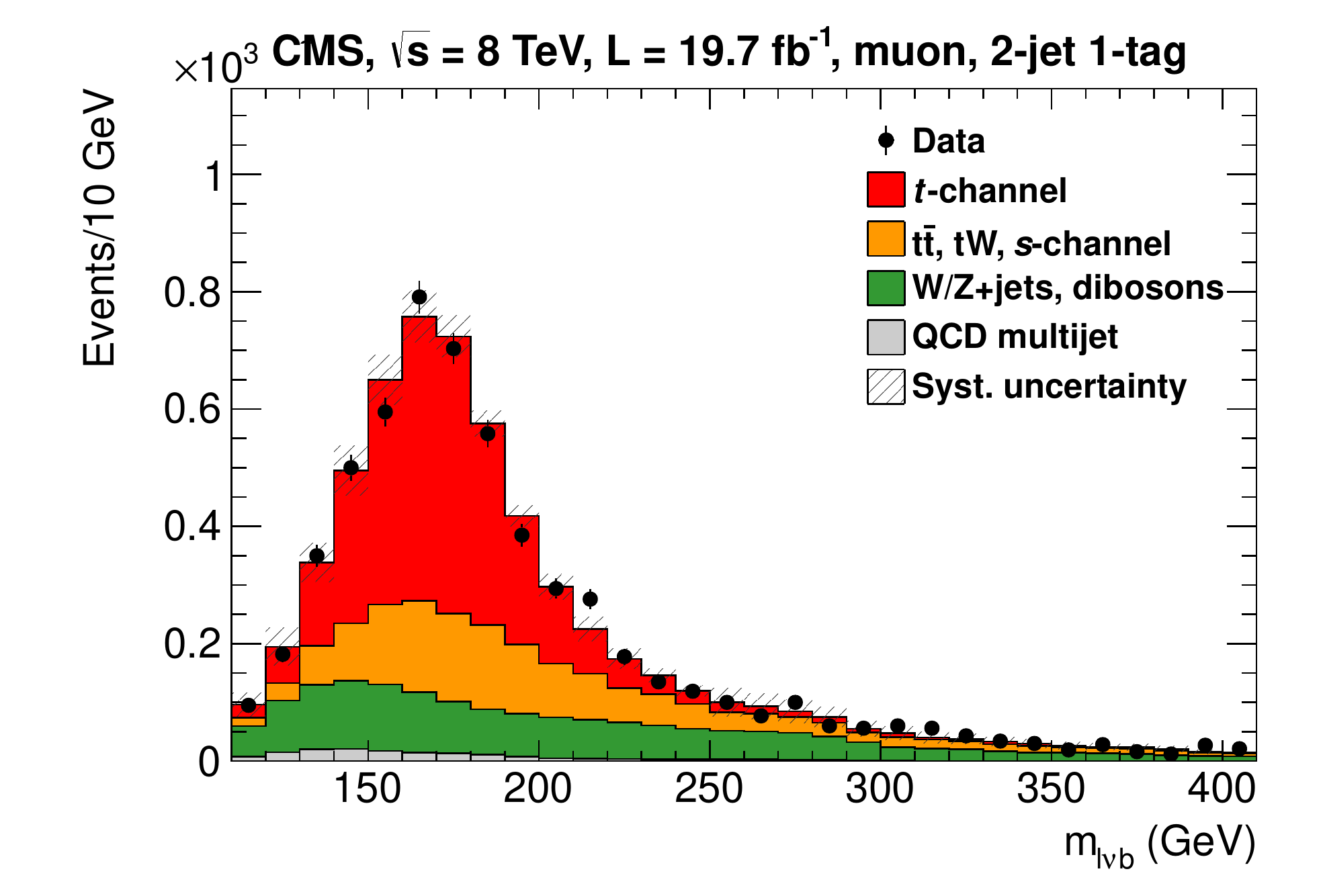}
\includegraphics[width=\cmsFigWidth]{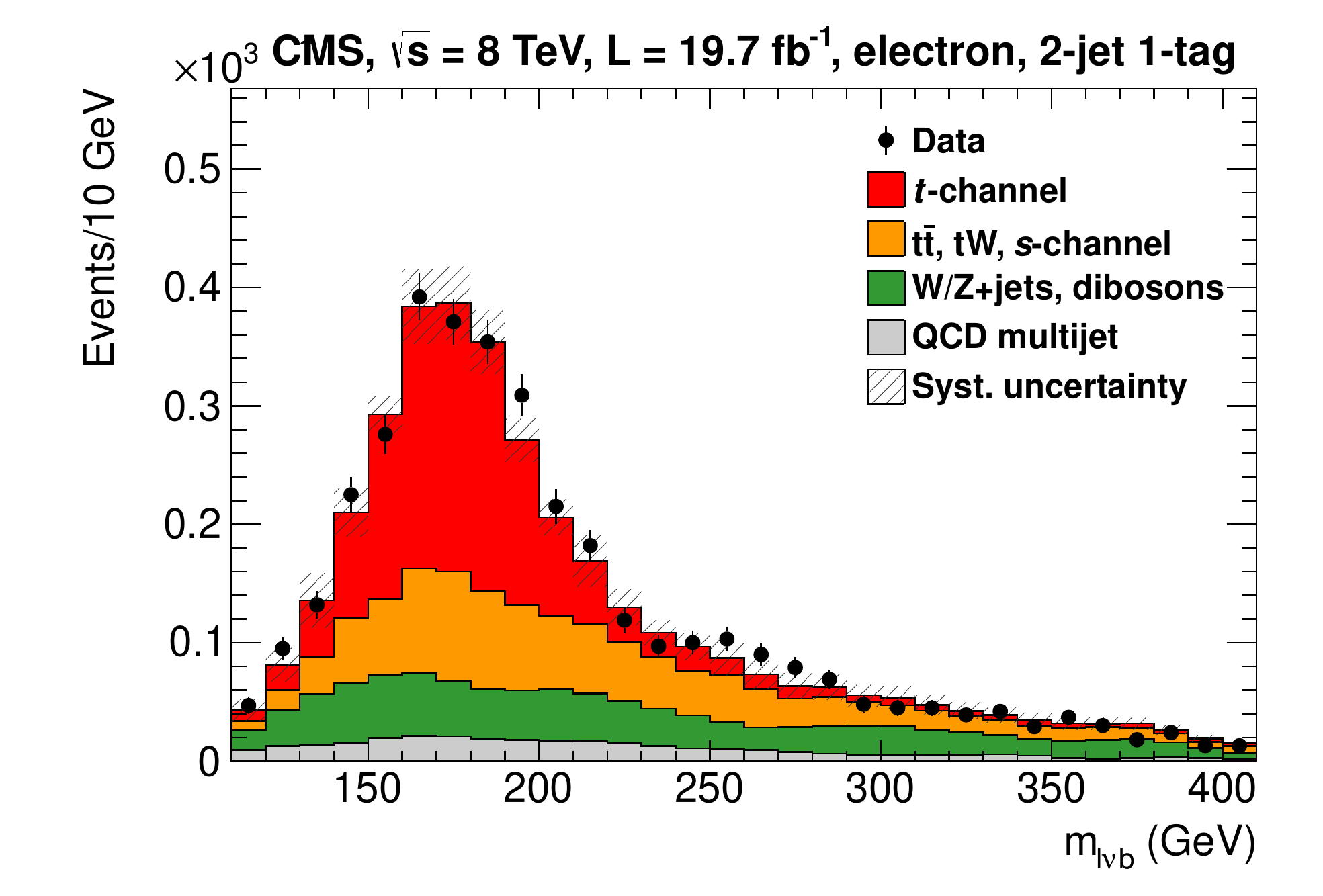}
\caption{\label{fig:fitsTopMass} Distribution of reconstructed top-quark mass \mt for muon (left) and electron (right) decay channels, in the region with $\absetalj > 2.5$, the contribution of each process is scaled to the cross section derived from the fit. Systematic uncertainty bands include the shape uncertainties on the distributions and uncertainties on the normalisation in the $\absetalj > 2.5$ region.}
\end{figure} 
\section{Systematic uncertainties}
\label{sec:syst}
\label{sec:systs}
\label{sec:systematics}
\label{sec:Systematics}
Contributions to the total systematic uncertainty are evaluated, with the exception of the uncertainties on the background estimation
described in section ~\ref{sec:backgrounds} and on the simulated samples size, with the following procedure:
pseudo-experiments are constructed using for each process the distributions and the yields generated considering the altered scenario.
A fit to the $\absetalj$ distribution is then performed for each pseudo-experiment with the nominal setup,
and the mean shift of the fit results with respect to the value obtained for the nominal fit is taken as the corresponding uncertainty.
A detailed description of each source of systematic uncertainty and of the treatment of uncertainties related to the data-based background estimation and to the size of simulated samples follows:
\begin{itemize}
\item \textbf{Jet energy scale (JES), jet energy resolution (JER), and missing transverse energy}:
All reconstructed jet four-momenta in simulated events are simultaneously varied according to the $\eta$- and $\pt$-dependent
uncertainties in the jet energy scale and resolution. The variation of jet momenta causes the
total momentum in the transverse plane to change, thus affecting the $\ETslash$ as well.
The component of the missing transverse energy that is not due to particles reconstructed as leptons and photons or clustered in
jets (``unclustered $\ETslash$'') is varied by $\pm$10\% ~\cite{Chatrchyan:2011ds}.
\item \textbf{Pileup}: The uncertainty in the average expected number of additional interactions per bunch crossing ($\pm$5\%)
is propagated as a systematic uncertainty to this measurement.
\item  \textbf{B-tagging}:  B-tagging and misidentification (mis-tag) efficiencies are estimated from control samples~\cite{CMS-PAS-BTV-11-004}. Scale factors are applied to simulated samples to reproduce efficiencies in data and
the corresponding uncertainties are propagated as systematic uncertainties.
\item \textbf{Muon/electron trigger and reconstruction}: Single-muon and single-electron trigger efficiency and reconstruction efficiency
as a function of the lepton $\eta$ and $\pt$  are estimated with a ``tag-and-probe'' method based on Drell--Yan data, as described in ref.~\cite{wzpaper}.
The effect of the incorrect determination of the muon charge is negligible, while for electrons the
uncertainty on the determination of the charge has been measured at $\sqrt{s}=7\TeV$ in ref.~\cite{EWK-10-006}.
\item \textbf{\wjets, \ttbar, and \QCD estimation}: The distributions and normalisations
of these three main backgrounds are derived mostly from data as described in section~\ref{sec:backgrounds}.
The uncertainty related to the \wjets and \ttbar\ estimation is
evaluated by generating pseudo-experiments in the SB and in the 3-jet 2-tag sample. The background estimation
is repeated, and then the fit to $\absetalj$ is performed and the uncertainty is taken as the \textsc{rms} of the
distribution of fit results.
An uncertainty in the \wjets contribution is obtained from alternative  $\absetalj$ shapes derived from simulation by varying the
W+b-jets and the W+c-jets background fractions by $\pm$30\% independently in the SR and SB regions.
An additional uncertainty in the $\ttbar$ estimation procedure is determined by performing the signal extraction using
the $\ttbar$ distribution in the entire \mt range, then using two different distributions for the signal and background regions.
The difference of the two results is taken as the uncertainty.
The $\QCD$ normalisation is varied by $\pm$50\% independently for muon and electron decay channels. This variation range
is obtained by performing the multijet estimation under different conditions and assumptions as described in section~\ref{sec:bkg},
and taking the maximum difference with respect to the value obtained with the nominal estimation procedure.
Additionally, all other systematic uncertainties are coherently propagated through the estimation procedure.
\item \textbf{Background normalisation}:
An uncertainty in the $\ttbar$ normalisation of $\pm$10\% is considered, covering the difference between theoretical predictions
in~\cite{Kidonakis:2012db} and~\cite{Cacciari:2011hy}.
For dibosons and single-top-quark tW and $s$-channel production the assumed uncertainty is $\pm$30\%, motivated by refs.~\cite{mcfm,Kidonakis:2012db}.
\item \textbf{Signal modelling}: Renormalisation and factorisation scales used in the signal simulation are varied by a factor 2 up and down,
and the corresponding variation is considered as the systematic uncertainty. The uncertainty on the simulation is obtained by comparing
the results obtained with the nominal {\POWHEG} signal samples with the ones obtained using samples generated by \textsc{ CompHEP}~\cite{comphep_1,comphep_2}.
Half of the difference is taken as systematic uncertainty.
\item \textbf{PDFs}: The uncertainty due to the choice of the PDF set is estimated by reweighting the simulated events
and repeating the signal extraction procedure. The envelope of the CT10~\cite{CT10}, MSTW~\cite{MSTW2008NLO}, and NNPDF~\cite{NNPDF} PDF sets
is taken as uncertainty, according to the PDF4LHC recommendations~\cite{PDF4LHC}.
\item \textbf{Simulation sample size:} The statistical uncertainty due to the limited size of simulated samples is taken into account by generating pseudo-experiments reproducing the statistical fluctuations of the model.
The fit procedure is repeated for each pseudo-experiment and the uncertainty is evaluated as the \textsc{rms} of
the distribution of fit results.
\item \textbf{Luminosity}: The integrated luminosity is known with a relative uncertainty of $\pm$2.6\%~\cite{lumi}.
\end{itemize}

The contribution of each source of uncertainty to the cross section and their ratio measurements is shown in tables ~\ref{tab:yieldmu-ele} and
~\ref{tab:yield_chmu-ele}, respectively.
Uncertainties due to the limited size of simulated and control samples in data for the background estimation do not cancel
and thus have an impact on the ratio measurement larger than on the total cross section.
Uncertainties that affect the signal efficiency in a similar way for single \tq and \tqbar, such as the b-tagging, or the lepton trigger
and reconstruction efficiencies, tend to cancel in the cross section ratio, thus have a smaller impact on its measurement. The luminosity uncertainty cancels as well in the ratio.
Uncertainties that affect the background processes that are independent from the lepton charge, like the \ttbar or the \QCD, have a bigger impact on
the single \tqbar cross section, for which the signal-to-background ratio is less favourable, and for this reason they do not cancel out entirely
in the ratio measurement.
Since single \tq and \tqbar production depend on different quark PDFs, the corresponding
PDF uncertainties are largely anticorrelated, and the corresponding contribution is enhanced in the charge ratio measurement.
As the momentum and pseudorapidity spectra of quarks and leptons for the single \tq and \tqbar processes are different,
the modelling uncertainties and the uncertainties from the jet energy scale and missing transverse energy do not fully cancel out in the ratio measurement.

Because of these differences, the event yields returned by the inclusive single-top-quark cross section and the single \tq and \tqbar cross section fits are not numerically identical. A consequence of this is that the values for the total cross section obtained in the two fits differ.
In particular the uncertainty in the heavy-flavour component is anticorrelated between the two measurements, and the
theoretical uncertainties tend to affect the exclusive extraction more than the inclusive one.

The choice to keep two separate procedures is motivated by the fact that the inclusive fit has
 a better overall performance regarding the systematic uncertainties in the inclusive cross section measurement.

 \begin{table}
 \topcaption{Relative impact of systematic uncertainties for the combined muon and electron decay channels.}
 \centering
 \begin{tabular}{ |c|c| }
 \hline
Uncertainty source  & $\sigma_{\tch}$ (\%) \\
\hline
 Statistical uncertainty & $\pm$ 2.7\\
\hline
 JES, JER, MET, and pileup  & $\pm$ 4.3\\
 b-tagging and mis-tag  & $\pm$ 2.5\\
 Lepton reconstruction/trig.  & $\pm$ 0.6\\
 \QCD estimation & $\pm$ 2.3\\
 \wjets, \ttbar estimation & $\pm$ 2.2\\
 Other backgrounds ratio & $\pm$ 0.3\\
 Signal modeling & $\pm$ 5.7\\
 PDF uncertainty & $\pm$ 1.9\\
 Simulation sample size & $\pm$ 0.7\\
 Luminosity & $\pm$ 2.6\\
 \hline
Total systematic  & $\pm$  8.9\\
 \hline
Total uncertainty  & $\pm$  9.3\\
 \hline
 \hline
Measured cross section & 83.6 $\pm$ 7.8\unit{pb} \\
 \hline
 \end{tabular}
 \label{tab:yieldmu-ele}
 \end{table}

 \begin{table}
\topcaption{Relative impact of systematic uncertainties on the exclusive single \tq and \tqbar
production cross sections and the ratio measurements.}
 \centering
 \begin{tabular}{ |c|c|c|c|}
 \hline
Uncertainty source  & $\sigma_{\tch}(\cPqt)$ (\%) & $\sigma_{\tch}(\cPaqt)$ (\%) & $R_{\tch}$ (\%) \\
\hline
 Statistical uncertainty & $\pm$ 2.7 & $\pm$ 4.9 & $\pm$ 5.1\\
\hline
 JES, JER, MET, and pileup  & $\pm$ 4.2 & $\pm$ 5.2 & $\pm$ 1.1\\
 b-tagging and mis-tag  & $\pm$ 2.6 & $\pm$ 2.6 & $\pm$ 0.2\\
 Lepton reconstruction/trig.  & $\pm$ 0.5 & $\pm$ 0.5 & $\pm$0.3\\
 \QCD estimation & $\pm$ 1.6 & $\pm$ 3.5 & $\pm$1.9\\
 \wjets, \ttbar estimation & $\pm$ 1.7 & $\pm$ 3.6 & $\pm$ 3.0\\
 Other backgrounds ratio & $\pm$ 0.1 & $\pm$ 0.2 & $\pm$ 0.6\\
 Signal modeling & $\pm$ 4.9 & $\pm$ 9.4 & $\pm$ 6.1\\
 PDF uncertainty & $\pm$ 2.5 & $\pm$ 4.8 & $\pm$ 6.2\\
 Simulation sample size & $\pm$ 0.6 & $\pm$ 1.1 & $\pm$ 1.2\\
 Luminosity & $\pm$ 2.6 & $\pm$ 2.6 & \text{---} \\
 \hline
Total systematic & $\pm$ 8.2 & $\pm$  13.4 & $\pm$  9.6\\
 \hline
Total uncertainty & $\pm$ 8.7 & $\pm$  14.2 & $\pm$  10.9\\
 \hline
 \hline
Measured cross section or ratio & 53.8 $\pm$ 4.7\unit{pb}  & 27.6 $\pm$ 3.9\unit{pb}  & 1.95 $\pm$ 0.21  \\
 \hline
 \end{tabular}
 \label{tab:yield_chmu-ele}
 \end{table}

\section{Results}
\label{sec:results}
\subsection{Cross section measurements}
The measured inclusive single-top-quark production cross section in the $t$-channel is
\begin{equation}
\sigma_{\tch} = \totmuelescorrxsec \pm \totmuelestatsscorrxsec\stat \pm \totmuelesystsscorrxsec\syst\unit{pb}.
\label{eq:inclusivexsec}
\end{equation}
The measured single \tq and \tqbar production cross sections in the $t$-channel are
\begin{equation}\begin{aligned}
\sigma_{\tch}(\cPqt) &= \chmuelespluscorrxsec \pm \chmuelestatsspluscorrxsec\stat \pm \chmuelesystsspluscorrxsec\syst\unit{pb},\\
\sigma_{\tch}(\cPaqt) &= \chmuelesminuscorrxsec  \pm \chmuelestatssminuscorrxsec\stat \pm \chmuelesystssminuscorrxsec\syst\unit{pb}.
\end{aligned}\end{equation}

A comparison of the currently available measurements of the inclusive cross section with the SM expectation obtained with a QCD computation at NLO with MCR in the 5F scheme~\cite{Campbell:2009gj} and at NLO+NNLL~\cite{Kidonakis:2011wy} is shown in figure~\ref{fig:finalplot}. The measurement is compared to the previous CMS $t$-channel cross section measurement at $\sqrt{s}=7\TeV$~\cite{Chatrchyan:2012vp} and the Betatron measurements at $\sqrt{s}=1.96\TeV$~\cite{D0-s,CDF-singletop}. The measurements are compared with the QCD expectations computed at NLO with MCR in the 5F scheme and at NLO+NNLL. The error band (width of the curve) is obtained by varying the top-quark mass within its current uncertainty~\cite{Group:2009qk}, estimating the PDF uncertainty according to the HEPDATA recommendations ~\cite{Campbell:2006wx}, and varying the factorisation and renormalisation scales coherently by a factor two up and down. The prediction in pp collisions can be also compared with the one at $\Pp\Pap$ because the inclusive single-top-quark cross section does not depend on whether the light quark originates from a proton or from an antiproton.

\begin{figure}[ht]
\centering
  \includegraphics[width=1.0\textwidth]{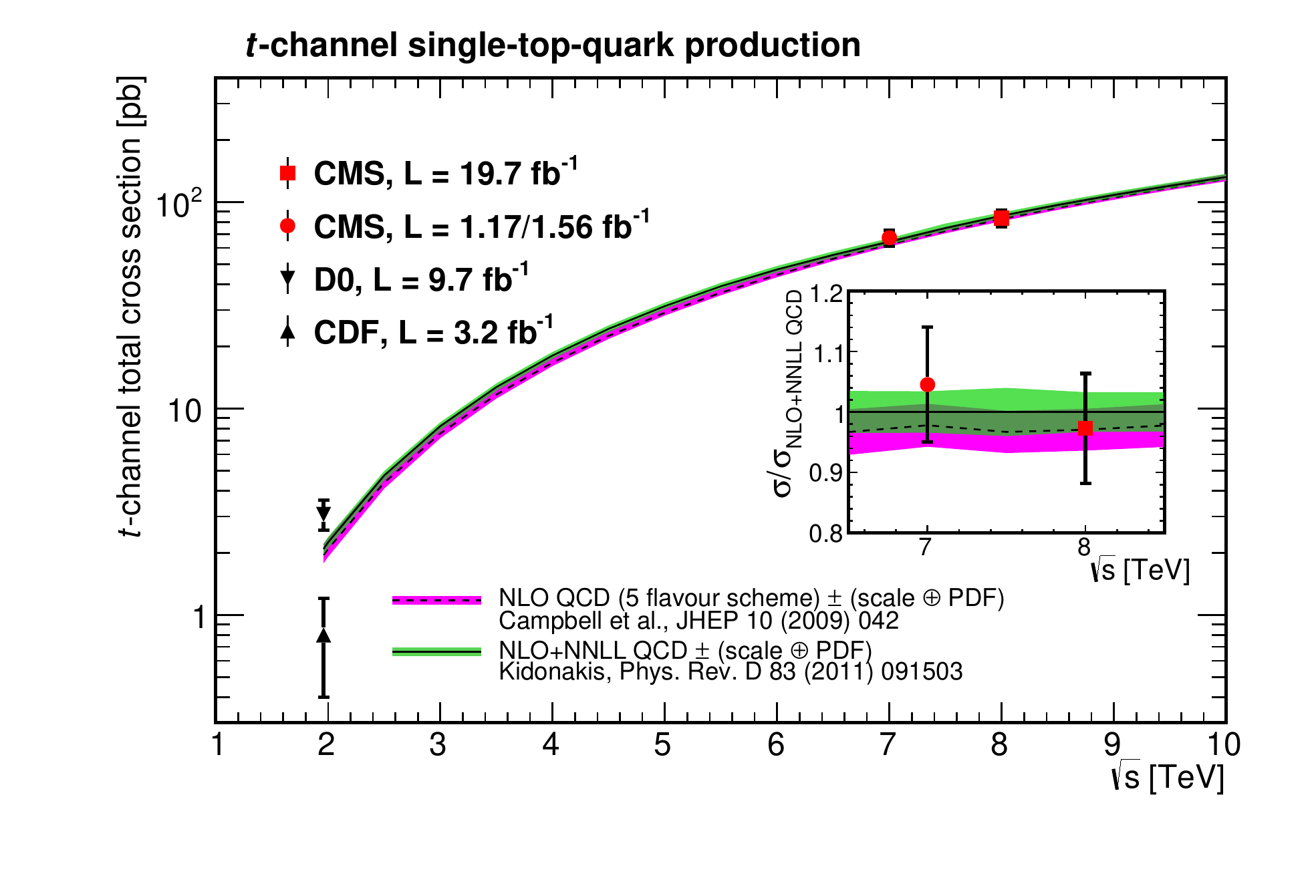}
  \caption{\label{fig:finalplot} Single-top-quark production cross section in the $t$-channel versus collider centre-of-mass energy.}
\end{figure}
\subsection{Cross section ratios}

The ratio of $t$-channel production cross sections at $\sqrt{s}=8$ and 7\TeV is derived with respect to the result reported in ref.~\cite{Chatrchyan:2012vp} for the single-top-quark $t$-channel cross section at $\sqrt{s}=7\TeV$. Three measurements are combined in ref.~\cite{Chatrchyan:2012vp}: two multivariate analyses and one, the $\etalj$ analysis, making use of a strategy and a selection that are close to the ones reported in this paper.
The correlations between the sources of uncertainties reported in section~\ref{sec:systs} and those in ref.~\cite{Chatrchyan:2012vp} are determined in the following way: the uncertainties related to signal extraction and background estimation from data are treated as fully uncorrelated between 7 and 8\TeV, while for the rest of the uncertainties the 8\TeV analysis is considered fully correlated with respect to its 7\TeV $\etalj$ counterpart, and the same choices for correlation as in~\cite{Chatrchyan:2012vp} are adopted between the 8\TeV $\etalj$ analysis and the two 7\TeV multivariate analyses. Taking into account the correlations as described, the measured ratio is

\begin{equation}
R_{8/7} = \sigma_{\tch}(8\TeV)/\sigma_{\tch}(7\TeV) =  \RatioSevenEight \pm \RatioSevenEightUncStat\stat \pm \RatioSevenEightUncSyst\,\mathrm{(syst.)}.
\end{equation}
The measured ratio of single \tq to \tqbar production cross sections at $\sqrt{s}=8$\TeV is
\begin{equation}
R_{\tch} = \sigma_{\tch}(\cPqt)/\sigma_{\tch}(\cPaqt) = \chmuelesratioRcorr \pm \chmuelestatssratioRcorr\stat \pm \chmuelesystssratioRcorr\syst.
\end{equation}
A comparison is shown in figure~\ref{fig:Ratio} of the measured $R_{\tch}$ to the predictions obtained with several PDF sets: MSTW2008NLO~\cite{MSTW2008NLO}, HERAPDF1.5 NLO~\cite{HERAPDF}, ABM11~\cite{ABM11}, CT10, CT10w~\cite{CT10}, and NNPDF~\cite{NNPDF}. For MSTW2008NLO, NNPDF, ABM, and CT10w the fixed 4F scheme PDFs are used together with the \POWHEG 4F scheme calculation. The \POWHEG calculation in the 5F scheme is used for all other PDFs, as they are derived from a variable flavour scheme. The nominal value for the top-quark mass used is 173.0\GeV. Error bars for the CMS measurement include the statistical (light yellow) and systematic (dark green) components. Error bars for the different PDF sets include the statistical uncertainty, the uncertainty in the factorisation and renormalisation scales, derived varying both of them by a factor 1/2 and 2, and the uncertainty in the top-quark mass, derived varying the top-quark mass between 172.0 and 174.0\GeV .
The different PDF sets predictions for this observable are not always compatible with each other within the respective uncertainties, thus displaying the potential for this measurement to discriminate between the different sets, should a better precision be achieved.
\begin{figure}[ht]
\centering
  \includegraphics[width=0.82\textwidth]{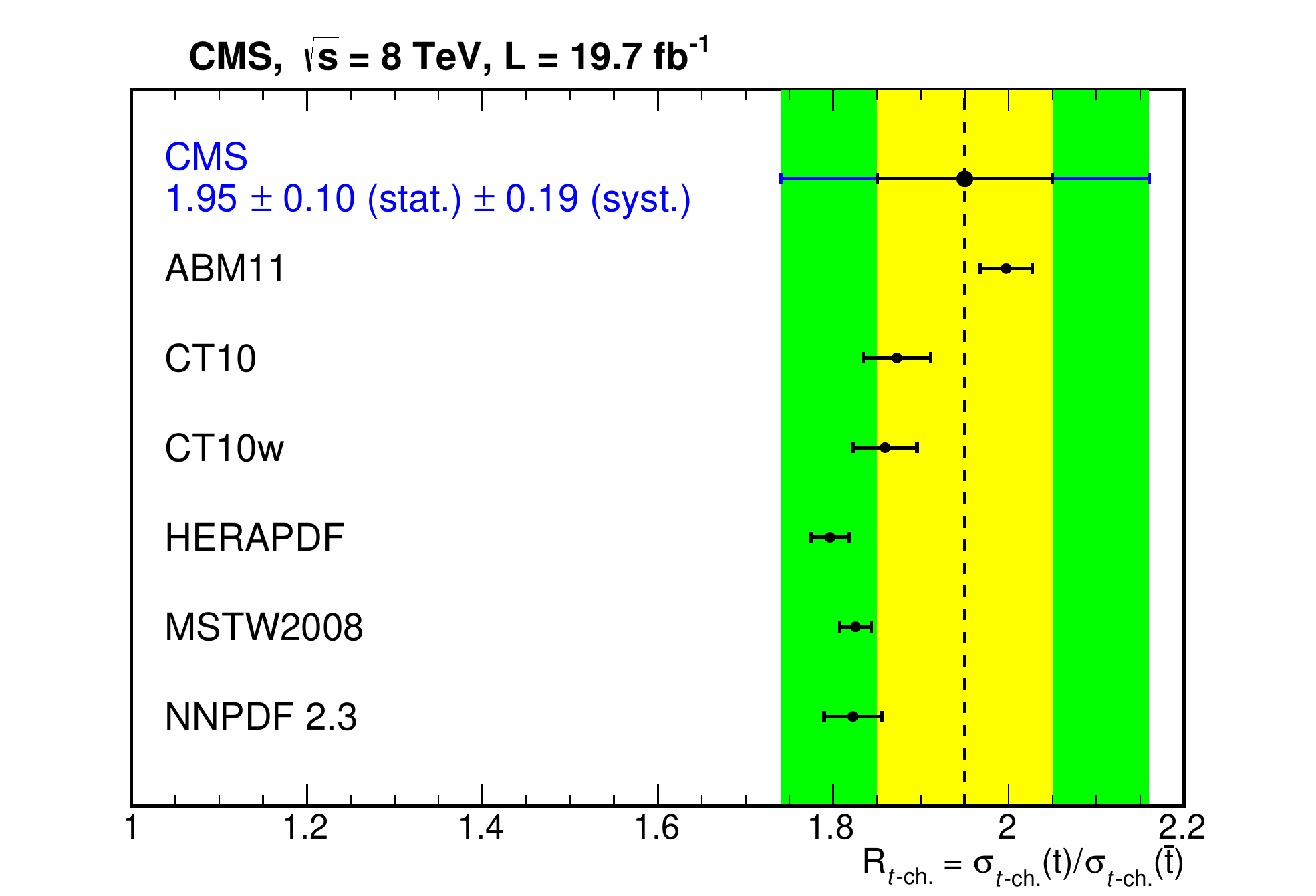}
\caption{\label{fig:Ratio} Comparison of the measured $R_{\tch}$ with the predictions obtained using different PDF sets.}
\end{figure}

\subsection{Extraction of \texorpdfstring{\absvtb}{the absolute value of Vtb}}

A feature of $t$-channel single-top-quark production is the presence of a Wtb vertex. This allows for an interpretation of
the cross section measurement in terms of the parameters regulating the strength of this coupling, most notably
the CKM matrix element $\vtb$.
The presence of anomalous couplings at the Wtb vertex can produce anomalous
form factors~\cite{Wtb:anom_1, Wtb:anom_2, Rizzo:1995uv} which are parametrised as $\flv$, where ``Lv'' refers to the specific left-handed vector nature of the couplings that
would modify the interaction strength.
In the approximation $\absvtd,\absvts \ll \absvtb$, we consider the top-quark decay branching fraction into Wb, $\mathcal{B}$, to be almost equal to 1, thus obtaining $\absflvvtb = \sqrt{\smash[b]{\ttot/\ttotth}}$. The choice of this approximation is motivated by the fact that several scenarios beyond the SM predict a deviation of the measured value of
$\flv$ from 1, but only a mild modification of $\mathcal{B}$~\cite{isvtbone}.
This allows to interpret a possible deviation from SM single-top-quark production cross section in terms of new physics. In the standard model case, $\flv = 1$, implying that the cross section measurement yields a direct constraint on $\absvtb$.
Thus inserting in the definition for $\absflvvtb$ the measured cross section from equation~\ref{eq:inclusivexsec} and the theoretical cross section from equation~\ref{eq:inclusivexsectheor}
results in
\begin{equation}
\absflvvtb = \VtbValue \pm \VtbValueUnc\exper\pm 0.016\thy,
\label{eq:vtb8teV}
\end{equation}
where both the experimental and the theoretical uncertainties are reported. The former comes from the uncertainties on the measurement of $\sigma_{\tch}$, while the latter comes from the uncertainties on $\ttotth$.
A similar measurement of \absflvvtb is performed in ref.~\cite{Chatrchyan:2012vp}. The results for \absflvvtb from this paper and from the three analyses in~\cite{Chatrchyan:2012vp} are combined using the best linear unbiased estimator (BLUE)~\cite{BLUE} method,
considering the full correlation matrix amongst the four measurements and the correlations described for the $R_{8/7}$ measurement, obtaining the following result:
\begin{equation}
\label{eq:vtbcomb}
\absflvvtb = \VtbValueComb \pm \VtbValueUncComb\exper\pm 0.016\thy \qquad\qquad \text{(7+8\TeV combination)}.
\end{equation}
This result can be directly compared with the current world average of $\absvtb$ from the Particle Data Group~\cite{pdg}, which is performed without the unitarity constraints on the CKM matrix and, using the above formalism for non-SM contributions, yields $\absflvvtb = 0.89\pm 0.07$.
From the result in equation~\ref{eq:vtbcomb}, the confidence interval for $\absVtb$, assuming the constraints $\absVtb \leq 1$ and $\flv=1$, is determined using the Feldman--Cousins unified approach~\cite{FC}, being $\absVtb > 0.92$ at the 95\% confidence level.

\section{Summary}
\label{sec:conclusion}
The total cross sections for production in the $t$-channel of single top quarks and individual single \tq and \tqbar have been measured in proton-proton collisions at the LHC at $\sqrt{s}$ = 8\TeV. The inclusive single-top-quark $t$-channel cross section has been measured to be $\sigma_{\tch} = \totmuelescorrxsec \pm \totmuelestatsscorrxsec\stat \pm \totmuelesystsscorrxsec\syst\unit{pb}$.
The single \tq and \tqbar cross sections have been measured to be $\sigma_{\tch}(\cPqt) = \chmuelespluscorrxsec \pm \chmuelestatsspluscorrxsec\stat\pm \chmuelesystsspluscorrxsec\syst\unit{pb}$ and $\sigma_{\tch}(\cPaqt) = \chmuelesminuscorrxsec \pm \chmuelestatssminuscorrxsec\stat\pm \chmuelesystssminuscorrxsec\syst\unit{pb}$, respectively. Their ratio has been found to be $R_{\tch}= \chmuelesratioRcorr\pm \chmuelestatssratioRcorr\stat \pm \chmuelesystssratioRcorr\syst$.
The ratio of $t$-channel single-top-quark production cross sections at $\sqrt{s}=8$ and 7 TeV has been measured to be
$R_{8/7} = \RatioSevenEight \pm \RatioSevenEightUncStat\stat\pm \RatioSevenEightUncSyst\syst$.
These measurements are in agreement with the standard model predictions.
From the measured single-top-quark production cross section, the modulus of the CKM matrix element $\Vtb$ has been determined. This result has been combined with the previous CMS measurement at 7 TeV, yielding the most precise measurement of its kind up to date:
$\absvtb = \VtbValueComb \pm  \VtbValueUncComb\exper \pm 0.016\thy$. Assuming $\absVtb \leq 1$, the 95\% confidence level limit has been found to be $\absVtb > 0.92$.

\section*{Acknowledgements}
We congratulate our colleagues in the CERN accelerator departments for the excellent performance of the LHC and thank the technical and administrative staffs at CERN and at other CMS institutes for their contributions to the success of the CMS effort. In addition, we gratefully acknowledge the computing centres and personnel of the Worldwide LHC Computing Grid for delivering so effectively the computing infrastructure essential to our analyses. Finally, we acknowledge the enduring support for the construction and operation of the LHC and the CMS detector provided by the following funding agencies: BMWF and FWF (Austria); FNRS and FWO (Belgium); CNPq, CAPES, FAPERJ, and FAPESP (Brazil); MES (Bulgaria); CERN; CAS, MoST, and NSFC (China); COLCIENCIAS (Colombia); MSES and CSF (Croatia); RPF (Cyprus); MoER, SF0690030s09 and ERDF (Estonia); Academy of Finland, MEC, and HIP (Finland); CEA and CNRS/IN2P3 (France); BMBF, DFG, and HGF (Germany); GSRT (Greece); OTKA and NIH (Hungary); DAE and DST (India); IPM (Iran); SFI (Ireland); INFN (Italy); NRF and WCU (Republic of Korea); LAS (Lithuania); MOE and UM (Malaysia); CINVESTAV, CONACYT, SEP, and UASLP-FAI (Mexico); MBIE (New Zealand); PAEC (Pakistan); MSHE and NSC (Poland); FCT (Portugal); JINR (Dubna); MON, RosAtom, RAS and RFBR (Russia); MESTD (Serbia); SEIDI and CPAN (Spain); Swiss Funding Agencies (Switzerland); NSC (Taipei); ThEPCenter, IPST, STAR and NSTDA (Thailand); TUBITAK and TAEK (Turkey); NASU and SFFR (Ukraine); STFC (United Kingdom); DOE and NSF (USA).

Individuals have received support from the Marie-Curie programme and the European Research Council and EPLANET (European Union); the Leventis Foundation; the A. P. Sloan Foundation; the Alexander von Humboldt Foundation; the Belgian Federal Science Policy Office; the Fonds pour la Formation \`a la Recherche dans l'Industrie et dans l'Agriculture (FRIA-Belgium); the Agentschap voor Innovatie door Wetenschap en Technologie (IWT-Belgium); the Ministry of Education, Youth and Sports (MEYS) of Czech Republic; the Council of Science and Industrial Research, India; the Compagnia di San Paolo (Torino); the HOMING PLUS programme of Foundation for Polish Science, cofinanced by EU, Regional Development Fund; and the Thalis and Aristeia programmes cofinanced by EU-ESF and the Greek NSRF.

\bibliography{auto_generated}   

\cleardoublepage \appendix\section{The CMS Collaboration \label{app:collab}}\begin{sloppypar}\hyphenpenalty=5000\widowpenalty=500\clubpenalty=5000\textbf{Yerevan Physics Institute,  Yerevan,  Armenia}\\*[0pt]
V.~Khachatryan, A.M.~Sirunyan, A.~Tumasyan
\vskip\cmsinstskip
\textbf{Institut f\"{u}r Hochenergiephysik der OeAW,  Wien,  Austria}\\*[0pt]
W.~Adam, T.~Bergauer, M.~Dragicevic, J.~Er\"{o}, C.~Fabjan\cmsAuthorMark{1}, M.~Friedl, R.~Fr\"{u}hwirth\cmsAuthorMark{1}, V.M.~Ghete, C.~Hartl, N.~H\"{o}rmann, J.~Hrubec, M.~Jeitler\cmsAuthorMark{1}, W.~Kiesenhofer, V.~Kn\"{u}nz, M.~Krammer\cmsAuthorMark{1}, I.~Kr\"{a}tschmer, D.~Liko, I.~Mikulec, D.~Rabady\cmsAuthorMark{2}, B.~Rahbaran, H.~Rohringer, R.~Sch\"{o}fbeck, J.~Strauss, A.~Taurok, W.~Treberer-Treberspurg, W.~Waltenberger, C.-E.~Wulz\cmsAuthorMark{1}
% ...
\vskip\cmsinstskip
% ...
1:~~Also at Vienna University of Technology, Vienna, Austria\\
2:~~Also at CERN, European Organization for Nuclear Research, Geneva, Switzerland\\
% ...

\end{sloppypar}
\end{document}